\documentclass[prb,amsfonts,showpacs,preprint]{revtex4}
\usepackage{hyperref,graphics,bm}
\begin{document}
\widetext
\title{
Effects of Crystal Structure and the On-Site Coulomb Interactions on
the Electronic and Magnetic Structure of Pyrochlores
$A_2$Mo$_2$O$_7$ ($A$$=$ Y, Gd, and Nd) }
\author{I. V. Solovyev}
\email[Electronic address: ]{igor.solovyev@aist.go.jp}
\affiliation{
Tokura Spin Superstructure Project, \\
ERATO Japan Science and Technology Corporation, \\
c/o National Institute of Advanced Industrial Science and Technology, \\
Central 4, 1-1-1 Higashi, Tsukuba, Ibaraki 305-8562, Japan
}
\date{\today}

\widetext
\begin{abstract}
Being motivated by recent experimental studies, we investigate magnetic
structures of the Mo pyrochlores $A_2$Mo$_2$O$_7$ ($A$$=$ Y, Nd, and Gd) and their
impact on the electronic properties.
The latter are closely related with the behavior of twelve Mo($t_{2g}$) bands,
located near the Fermi level and well separated from the rest of the spectrum.
We use a
mean-field Hartree-Fock approach, which combines
fine details of the electronic structure for these bands,
extracted from the conventional
calculations in the local-density approximation, the spin-orbit interaction, and
the on-site Coulomb interactions amongst the Mo($4d$) electrons, treated
in the most general rotationally invariant form. The Coulomb repulsion $U$ plays a very important
role in the problem, and the semi-empirical value $U$$\sim$$1.5$-$2.5$ eV accounts
simultaneously for the metal-insulator (M-I) transition, the ferromagnetic (FM) -
spin-glass (SG) transition, and for the observed enhancement of the anomalous Hall effect (AHE).
The M-I transition is mainly controlled by $U$. The magnetic structure at the metallic
side is nearly collinear FM, due to the double exchange mechanism. The transition
into the insulating state is accompanied by the large canting of spin and orbital
magnetic moments. The sign of exchange interactions in the insulating state is
controlled by the Mo-Mo distances. Smaller distances favor the antiferromagnetic
coupling, which preludes the SG behavior in the frustrated pyrochlore lattice.
Large AHE is expected in the nearly collinear FM state, near the point of M-I transition,
and is related with the unquenched orbital magnetization at the Mo sites. We also
predict large magneto-optical effect in the same FM compounds.
\end{abstract}



\maketitle

\section{\label{sec:intr}Introduction}

  The pyrochlores with the chemical formula $A_2$Mo$_2$O$_7$ ($A$ being the divalent
element) exhibit a number of interesting and not completely understood phenomena,\cite{Taguchi3}
which present the main motivation for the analysis of electronic and magnetic structure
of these compounds, which we undertake
in this work.

\begin{figure}
\centering \noindent
\resizebox{10cm}{!}{\includegraphics{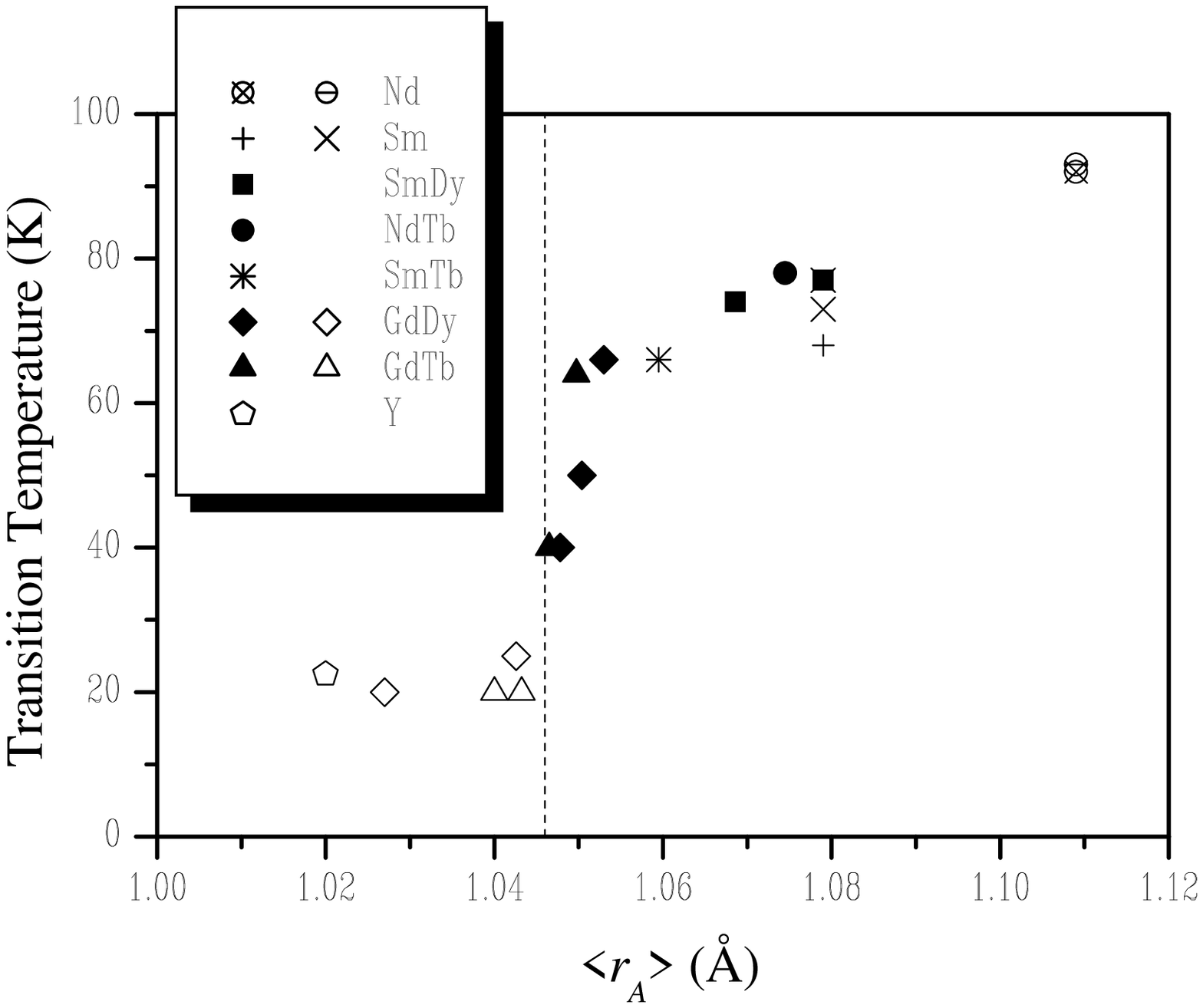}}
\caption{Experimental phase diagram of $A_2$Mo$_2$O$_7$: transition
temperature against the averaged ionic radius.
The values for Nd$_2$Mo$_2$O$_7$ are taken from Refs.~\protect\onlinecite{Moritomo}
($\otimes$) and \protect\onlinecite{Yoshii} ($\ominus$), for
Sm$_2$Mo$_2$O$_7$ -- from Refs.~\protect\onlinecite{Moritomo} ($+$) and
\protect\onlinecite{Taguchi} ($\times$), for
(Sm$_{1-x}$Dy$_x$)$_2$Mo$_2$O$_7$ ($x$$=$ 0 and 0.2),
(Nd$_{1-x}$Tb$_x$)$_2$Mo$_2$O$_7$ ($x$$=$ 0.5),
(Sm$_{1-x}$Tb$_x$)$_2$Mo$_2$O$_7$ ($x$$=$ 0.5),
(Gd$_{1-x}$Dy$_x$)$_2$Mo$_2$O$_7$ ($x$$=$ 0, 0.1, 0.2, 0.4, and 1),
and (Gd$_{1-x}$Tb$_x$)$_2$Mo$_2$O$_7$ ($x$$=$ 0.25, 0.5, 0.75, and 1)
-- from Ref~\protect\onlinecite{Katsufuji}, and for Y$_2$Mo$_2$O$_7$ -- from
Ref.~\protect\onlinecite{Gardner}.
Dotted line shows the boundary between normal ferromagnetic
(right) and spin-glass (left) behavior.}
\label{fig.Tcexp}
\end{figure}

  $\bullet$ The pyrochlores $A_2$Mo$_2$O$_7$ have rather puzzling magnetic phase diagram as a function
of averaged ionic radius of $A$-sites, $\langle r_A \rangle$
(Fig.~\ref{fig.Tcexp}).\cite{Katsufuji,Moritomo,Taguchi2,Iikubo}
Large $\langle r_A \rangle$ ($>$$R_c$$\sim$$1.047$) stabilizes the ferromagnetic (FM)
ground state. Typical examples of the FM pyrochlores are Nd$_2$Mo$_2$O$_7$ and
Gd$_2$Mo$_2$O$_7$. The Curie temperature ($T_C$) is of the order of 80 K and slowly increases
with $\langle r_A \rangle$. Smaller $\langle r_A \rangle$ ($<$$R_c$) give rise to
the spin-glass (SG) behavior. The characteristic transition temperature into the SG state
is of the order of 20 K.
The canonical example of the SG compounds is Y$_2$Mo$_2$O$_7$.\cite{Reimers,Gardner}
Now it is commonly accepted that the exchange coupling between
nearest-neighbor ({\it nn}) Mo spins changes the sign around $R_c$ and becomes
antiferromagnetic (AFM) in the SG region.
The pyrochlores with AFM interactions in the lattice present a typical example of
geometrically frustrated systems with infinitely degenerate magnetic ground state.\cite{Reimers2}
According to the recent experimental data,\cite{Booth,Keren} the SG behavior itself is
caused by the combination of the geometrical frustrations and the disorder of the local
lattice distortions. The latter is required in order to produce an inhomogeneity in the
distribution of the {\it nn} interactions and freeze the random spin configuration.
The origin of this behavior itself is very complicated, and
we will not be able to address it properly in our work. However, our main
concern will be in some sense more general: we will try to understand which parameter
of the crystal structure controls the sign of the {\it nn} magnetic interactions, and which
part of the electronic structure is responsible for the FM and AFM interactions
in these compounds? Contrary to the widespread point of view that the magnetic
ground state is controlled by the Mo-O-Mo angle, we will argue that the key parameter
is in fact the Mo-Mo distance, which is directly related with the unit cell volume
and controls the strength of direct interactions between extended Mo($4d$) orbitals.

  According to the experimental data, all Mo pyrochlores which show the SG behavior are
small-gap insulators. However, an opposite statement is generally incorrect and in
different compounds the ferromagnetism is known to coexist with the metallic as well as
insulating behavior.\cite{Iikubo} We will show that this trend can be naturally understood
in terms of the Mo-Mo distances and the strength of the on-site Coulomb interaction $U$
amongst the Mo($4d$) electrons.

  $\bullet$ The spin-orbit interaction (SOI) in the pyrochlore lattice should generally
lead to a non-collinear magnetic ordering associated with the modulation of
single-ion anisotropy axes.\cite{Moriya} The magnitude and the main source of this
non-collinearity (whether it is primarily associated with the Mo or $A$ sublattices,
and whether it reflects the ordering of spin or orbital magnetic moments) is largely
unresolved problem. In the present work we will try to rationalize this question by
focusing on the magnetic structure of the Mo sublattice. We will show that as long as
the system is metallic (that by itself depends on the strength of the Coulomb
interaction $U$), the non-collinearity is very small. However, the transition into
the insulating state is accompanied by the abrupt change of the spin and orbital
magnetic structures, which become essentially non-collinear.

  $\bullet$ Some ferromagnetic (or nearly ferromagnetic) pyrochlores exhibit rather
unusual anomalous Hall effect (AHE). A non-vanishing Hall conductivity ($\sigma_H$)
emerges near $T_C$, then monotonously increases with the decrease of the temperature $T$,
and reaches the maximal value ($\sigma_H$$\sim$$20$ $\Omega^{-1}{\rm cm}^{-1}$ for
Nd$_2$Mo$_2$O$_7$) near $T$$=$$0$. Both magnitude and the temperature dependence of the AHE
is in a drastic contrast with other oxide materials, such as colossal-magnetoresistive
manganites, for example.\cite{AHE_manganites}

  The theory of the AHE (originally called as the extraordinary Hall effect)
has very long and rather controversial history.\cite{KL,AHE_history}
Many controversies were related
with the lack of the general transport theory and partly --
with the lack of good computer facilities
in 1950s and 1960s, so that many discussions went around very simplified and not always
justified approximations, such as the free-electron and weak scattering limit,
the single-orbital tight-binding approach, the
classical Boltzmann transport equation, etc.\cite{comment.14}
The modern way to address the problem is to use
the Kubo formalism and relate $\sigma_H$ with the real part of the antisymmetric off-diagonal
element of the conductivity tensor.\cite{Kondorskii}
From this point of view there are no rigid constraints which
would forbid the existence of $\sigma_H$, even for periodic systems,
once the time-reversal symmetry is violated macroscopically.\cite{comment.10}

  The main question is which mechanism leads to the finite value of $\sigma_H$.
The first one was proposed by Karplus and Luttinger,\cite{KL}
who stressed the importance of SOI in the problem. Although the concrete scenario
considered by Karplus and Luttinger was not correct,\cite{comment.14}
they were certainly right by arguing that the
SOI is essential in order to couple
the spin polarization of the conduction electrons to the lattice and produce the
right-left asymmetry of the stationary Bloch states, which may lead to the
Hall current, perpendicular to both
the electric field and the magnetization.

  However, an exceptionally large $\sigma_H$ observed in the pyrochlore compounds urged
several authors to search for a new and unconventional mechanism for the AHE. Such
mechanism was proposed to be due to the spin chirality (basically, the solid angle
subtended by three non-coplanar spins) in essentially non-collinear magnetic
structure.\cite{Taguchi3,Nagaosa} The chirality acts as an effective field, which in
the combination with the peculiar lattice geometry is
believed to yield a finite $\sigma_H$.
Although the mechanism itself is rather exotic and there is a great deal of
cancellations between different elements of the non-collinear
spin structure,\cite{Dzyaloshinskii} these authors were certainly right by
stressing the crucial role of the inter-band transitions in the problem of AHE.

  In this work we will go back to a more conventional picture and argue that the
large value of $\sigma_H$ in some Mo pyrochlores can be explained by rather
unique combination of three factors: (i) large SOI associated with the Mo($4d$)
states in the nearly collinear FM structure; (ii) strong enhancement of the
SOI by the on-site Coulomb interactions in the $t_{2g}$ electron systems,\cite{PRL98}
and (iii) half-metallic electronic structure of the FM pyrochlores.\cite{Irkhin}

  Regarding the Coulomb interaction itself, which in our work will be treated as
a parameter, we will be able to present a very consistent picture by arguing that
$U$$\sim$$1.5$-$2.5$ eV in Mo pyrochlores accounts simultaneously for
(i) metal-insulator transition; (ii) FM-SG transition; (iii) required enhancement
for the AHE in the FM pyrochlore compounds.

  The rest of the paper is organized as follows. In Sec.~\ref{sec:main_details}
we remind the basic details of the crystal and electronic structure of
$A_2$Mo$_2$O$_7$. We consider three characteristic compounds corresponding to
$A$$=$ Y, Gd and Nd.
In Sec.~\ref{sec:model} we describe the model
Hartree-Fock (HF)
approach, which
combines the ideas of {\it ab-initio} electronic structure calculations with
the physics of degenerate Hubbard model on the pyrochlore
lattice. Secs.~\ref{sec:DOS} and
\ref{sec:exchange} summarize results of calculations without SOI
(correspondingly, the basic changes of the electronic structure and the orbital
ordering, and the behavior of inter-atomic magnetic interactions).
Secs.~\ref{sec:NC} and \ref{sec:Hall} are devoted to the analysis of
SOI related properties
(the non-collinear magnetic structure, the Hall conductivity, and the
magneto-optical effect). Finally, in Sec.~\ref{sec:summary} we discuss some subtle
points related with the interpretation of experimental data and give the
summary of the entire work.

\section{\label{sec:main_details}Main Details of Crystal and Electronic Structure}

  The pyrochlores $A_2$Mo$_2$O$_7$
crystallize in a face-centered cubic
structure with the space group $Fm\overline{3}d$, in which $A$ and Mo
occupy correspondingly $16d$ and $16c$ positions, and form interpenetrating
sublattices of the corner-sharing tetrahedra. There are two non-equivalent types of
oxygen sites, and only one internal parameter which may control the properties
of $A_2$Mo$_2$O$_7$. That is the coordinate $u$ of the O $48f$ sites.

\begin{figure}
\centering \noindent
\resizebox{7cm}{!}{\includegraphics{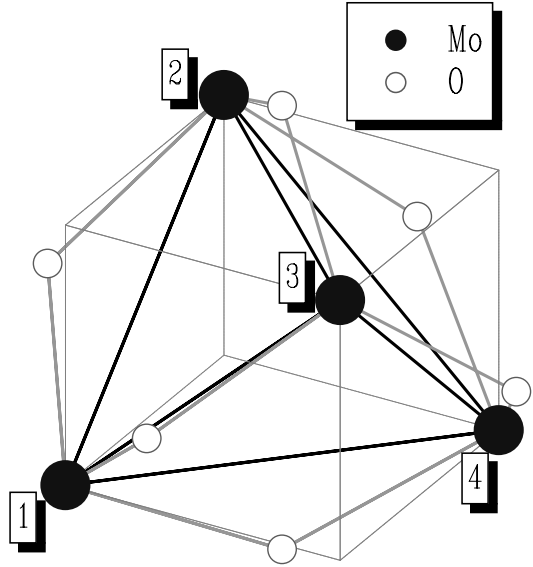}}
\caption{Relative position of the Mo and O $48f$ sites
in pyrochlore lattice.}
\label{fig.tetrahedron}
\end{figure}

  The single Mo tetrahedron is shown in Fig.~\ref{fig.tetrahedron}.
Four Mo sites are located at
$\bm{\tau}_1$$=$$(0,0,0)$,
$\bm{\tau}_2$$=$$(0,\frac{1}{4},\frac{1}{4})$,
$\bm{\tau}_3$$=$$(\frac{1}{4},0,\frac{1}{4})$,
and $\bm{\tau}_4$$=$$(\frac{1}{4},\frac{1}{4},0)$,
in units of cubic lattice parameter $a$.
Each Mo site has sixfold O $48f$ coordination. The oxygen atoms specify the local
coordinate frame around each Mo site. Around the site 1, it is given by
\begin{equation}
\mathbb{R}_1^{\alpha \beta} = \frac{1+(1-8u)\delta_{\alpha \beta}}{\sqrt{64u^2-32u+6}},
\label{eqn:lframe}
\end{equation}
where $\alpha ,\beta$$=$ $x$, $y$, or $z$.
$u$$=$$\frac{5}{16}$ corresponds to the perfect octahedral environment. In this case
$\widehat{\mathbb{R}}_1$$=$$\| \mathbb{R}_1^{\alpha \beta} \|$
is the standard $60^\circ$ rotation
around the cubic $[1,1,1]$ axis. $u$$>$$\frac{5}{16}$ gives rise to an additional
trigonal contraction of the local coordinate frame. Similar matrices associated with the sites
2, 3, and 4 can be obtained by the $180^\circ$ rotations of
$\widehat{\mathbb{R}}_1$ around
$x$, $y$, and $z$, respectively.

\begin{figure}
\centering \noindent
\resizebox{7cm}{!}{\includegraphics{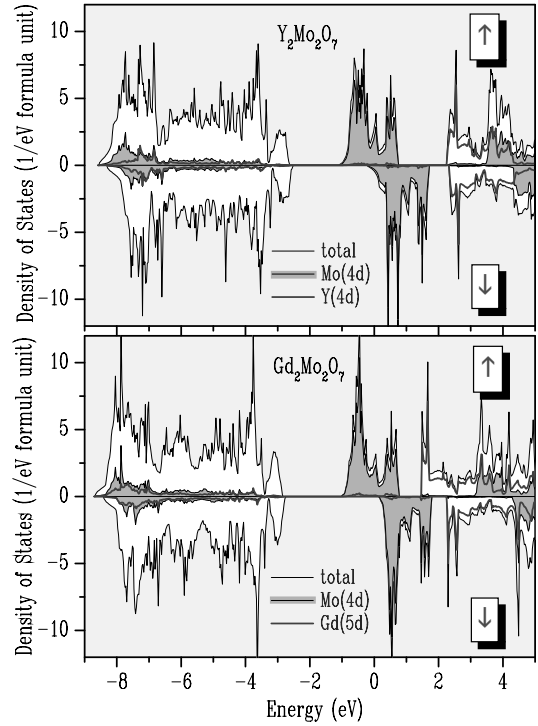}}
\caption{Total and partial densities of states of Y$_2$Mo$_2$O$_7$
and Gd$_2$Mo$_2$O$_7$
in the local-spin-density approximation. The Mo($t_{2g}$) states are
located near the Fermi level (chosen as zero of energy). The Mo($e_g$)
states emerge
around 4 eV.}
\label{fig.DOSLSDA}
\end{figure}

\begin{figure}
\centering \noindent
\resizebox{10cm}{!}{\includegraphics{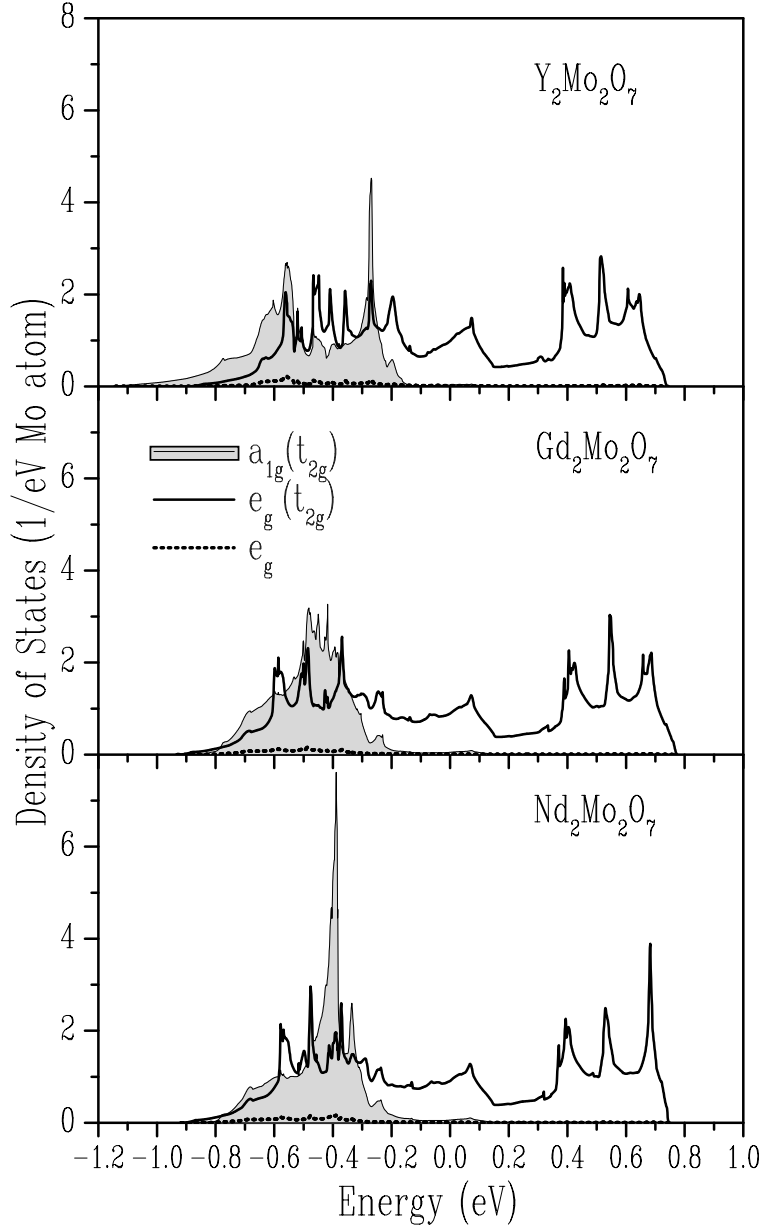}}
\caption{Mo($t_{2g}$) states in the local coordinate frame, split into the
one-dimensional $a_{1g}$ and two-dimensional $e_g'$ (denoted as
$e_g$($t_{2g}$)) representations by
the local trigonal distortion. The Fermi level is at zero.
The contribution of proper $e_g$ states to this region is small.}
\label{fig.DOSt2g}
\end{figure}

  Structural parameters of $A_2$Mo$_2$O$_7$ are listed in Table~\ref{tab:structure}.
Corresponding densities of states, obtained in the local-spin-density approximation
(LSDA) are shown in Figs.~\ref{fig.DOSLSDA} and \ref{fig.DOSt2g}.

  In the local coordinate frame, the Mo($4d$) orbitals are split into the triply-degenerate
$t_{2g}$ and doubly-degenerate $e_g$ states.
The splitting is of the order of 4 eV.
Twelve $t_{2g}$ bands
are located near the
Fermi level and well separated from the rest of the spectrum, which consists
of the broad O($2p$)
band spreading from -8.5 to -2.5 eV and either Y($4d$) or Gd/Nd($5d$) bands located
just above the group of the $t_{2g}$ states.
The Mo($e_g$) states are located in the higher part of the spectrum.
Interestingly enough that
all three compounds are ferromagnetic, even in the
LSDA, that is rather unusual for the $4d$ oxides, perhaps except the well know
example of SrRuO$_3$.\cite{Singh}

  The trigonal distortion
and different hybridization with the O($2p$) states
will further split the Mo($t_{2g}$) states into the
one-dimensional $a_{1g}$ and two-dimensional $e_g'$ representation.\cite{comment.3}

  The crystal structure affects the Mo($t_{2g}$) bandwidth via two mechanisms
(see Table~\ref{tab:structure}).
\begin{enumerate}
\item
The Mo-O-Mo angle, which decreases in the direction Nd$\rightarrow$Gd$\rightarrow$Y.
Therefore,
the interactions between Mo($t_{2g}$) orbitals which are mediated by the
O($2p$) states will also decrease. This effect is partly compensated by the
decrease of the Mo-O bondlength for $A$$=$ Gd and Y.
\item
The lattice parameter $a$ and the Mo-Mo distance, which decrease in the
direction Nd$\rightarrow$Gd$\rightarrow$Y by 2.6\%.
This will increase direct
interactions between extended Mo($4d$) orbitals.
\end{enumerate}

  Generally, these two effects act in the opposite directions and partly
compensate each other. For example, the width of the $e_g'$ band is practically
the same for all three compounds (Fig.~\ref{fig.DOSt2g}). On the other hand, the
$a_{1g}$ orbitals, whose lobes are the most distant from all neighboring oxygen
sites are mainly affected by the second mechanism, and the $a_{1g}$
bandwidth will {\it increase} in the direction Nd$\rightarrow$Gd$\rightarrow$Y.
As we will show in Sec.~\ref{sec:exchange}, this effect plays a crucial role
in the stabilization of AFM interactions in Y$_2$Mo$_2$O$_7$,
which preludes the SG behavior.

  Thus, despite an apparent complexity of the crystal structure,
the pyrochlores $A_2$Mo$_2$O$_7$ present rather simple example of the
electronic structure and in order to understand
the nature of fascinating electronic and magnetic properties of these compounds,
we need to concentrate on the behavior of twelve well isolated
Mo($t_{2g}$) bands.\cite{LiV2O4}

\section{\label{sec:model}
Model Hartree-Fock Approach}

  In this and subsequent sections we will further elaborate this picture, by cutting
the Mo($t_{2g}$) bands from the rest of the spectrum,
including the on-site
Coulomb interactions amongst the
Mo($4d$) electrons, and solving this problem
on the level of HF approximation. Hence, our model Hamiltonian
has the following form, in the basis of non-relativistic
LDA eigenfunctions $|n {\bf k}\rangle$ (LDA is the spin-restricted version of LSDA):
\begin{equation}
{\cal H}_{nn'}({\bf k})=\varepsilon_{n{\bf k}}\delta_{nn'} +
{\cal H}^{SOI}_{nn'}({\bf k}) + {\cal H}^{U}_{nn'}({\bf k}).
\label{eqn:Hamiltonian}
\end{equation}
The spin indices are already included to $n$ and $n'$, which
are running from $1$ to $12$$\times$$2$.
The first term in ${\cal H}_{nn'}({\bf k})$ is the
non-magnetic LDA part and
$\varepsilon_{n{\bf k}}$ are the LDA eigenvalues. ${\cal H}^{SOI}_{nn'}({\bf k})$ is
the matrix of relativistic SOI, which is computed in the
conventional way, ${\cal H}^{SOI}_{nn'}({\bf k})$$=$$
\langle n {\bf k}|\frac{1}{2} \xi({\bf r}) \left( {\bf L}, \bm{\sigma} \right)
|n' {\bf k}\rangle$,
where $\xi({\bf r})$$=$$\frac{1}{c^2} \frac{dV({\bf r})}{dr}$, ${\bf L}$ is the
operator of angular momentum,
$\bm{\sigma}$ is the vector of Pauli matrices,
$V({\bf r})$ is the self-consistent LDA potential, and $c$ is the velocity of
light ($c$$\approx$$274$ in Ry units). The inclusion of
${\cal H}^{SOI}_{nn'}({\bf k})$ into the model Hamiltonian is optional, that will be
always specified in the text.

  All magnetic effects are caused by the on-site
Coulomb interactions, ${\cal H}^{U}_{nn'}({\bf k})$, which can be written
in terms of projections onto an orthonormal set of atomic-like Mo($4d$) orbitals
$|{\bf R} \bm{\tau} \mu \rangle$:
$$
{\cal H}^{U}_{nn'}({\bf k}) = \sum_{{\bf R} \bm{\tau}} \sum_{\mu_1 \mu_2}
\langle n{\bf k}|{\bf R} \bm{\tau} \mu_1 \rangle
{\cal U}_{\bm{\tau}}^{\mu_1 \mu_2}
\langle {\bf R} \bm{\tau} \mu_2 | n' {\bf k} \rangle.
$$
where ${\bf R}$ are the translation vectors in the face-centered cubic
pyrochlore lattice, and $\bm{\tau}$ specify different
Mo sites in the unit cell.
$\mu$$\equiv$$\{s,m\}$ are the joint indices including the spin
($s$$=$ $\uparrow$ or
$\downarrow$) and orbital ($m$$=$ $xy$, $yz$, $3z^2$$-$$r^2$, $zx$, or
$x^2$$-$$y^2$) degrees of freedom. Matrix elements of the HF potential,
${\cal U}_{\bm{\tau}}^{\mu_1 \mu_2}$, can be expressed
in the most general rotationally-invariant form as:\cite{PRL98}
\begin{equation}
{\cal U}_{\bm{\tau}}^{\mu_1 \mu_2} =
\sum_{\mu_3 \mu_4} \left( U_{\mu_1 \mu_3 \mu_2 \mu_4} -
U_{\mu_1 \mu_3 \mu_4 \mu_2} \right) n_{\bm{\tau}}^{\mu_3 \mu_4},
\label{eqn:HFpotential}
\end{equation}
where $U_{\mu_1 \mu_3 \mu_2 \mu_4}$$=$$
\langle m_1 m_3 | \frac{1}{r_{12}} | m_2 m_4 \rangle \delta_{s_1 s_2} \delta_{s_3 s_4}$
are the matrix elements of the Coulomb interactions.
The latter are fully specified by
three radial Slater integrals:
$F^0$, $F^2$, and $F^4$, which are assumed to be {\it renormalized}
from the bare atomic values. Equivalently, one can introduce parameters
of the averaged Coulomb interaction, $U$$=$$F^0$,
the intra-atomic exchange coupling, $J$$=$$\frac{1}{14}(F^2$$+$$F^4)$, and the
non-sphericity of the Coulomb interactions between orbitals with the
same spin, $B$$=$$\frac{1}{441}(9F^2$$-$$5F^4)$.
Generally, both $U_{\rm eff}$$=$$U$$-$$J$
and $B$ contribute to the orbital polarization in solids and affect the properties
caused by violation of the time-inversion symmetry.\cite{PRL98}

  The density matrix $\widehat{n}_{\bm{\tau}}$$\equiv$$
\| n_{\bm{\tau}}^{\mu \mu'} \|$ can be obtained from
the Green function
$\widehat{\cal G}_{\bm{\tau} \bm{\tau}'}$$\equiv$$
\| {\cal G}_{\bm{\tau} \bm{\tau}'}^{\mu \mu'} \|$, using the
standard relation
\begin{equation}
\widehat{n}_{\bm{\tau}} = -\frac{1}{\pi} {\rm Im}
\int_{-\infty}^{\varepsilon_F} d \varepsilon
\widehat{\cal G}_{\bm{\tau} \bm{\tau}}(\varepsilon),
\label{eqn:dmatrix}
\end{equation}
where
\begin{equation}
{\cal G}_{\bm{\tau} \bm{\tau}'}^{\mu \mu'}(\varepsilon) =
\sum_n
\frac{1}{\Omega_{\rm BZ}}
\int d {\bf k}
\frac{\langle {\bf R} \bm{\tau} \mu | \widetilde{n {\bf k}} \rangle
\langle \widetilde{n {\bf k}} | {\bf R} \bm{\tau}' \mu' \rangle }
{\varepsilon - \tilde{\varepsilon}_{n{\bf k}} + i\delta}
\label{eqn:Green}
\end{equation}
is the spectral representation for the Green function
in terms of the eigenfunctions $| \widetilde{n {\bf k}} \rangle$ and the eigenvalues
$\tilde{\varepsilon}_{n{\bf k}}$ of the Hamiltonian (\ref{eqn:Hamiltonian}).
$\Omega_{\rm BZ}$ is the volume of the first Brillouin zone.

  Eqs.~(\ref{eqn:Hamiltonian}), (\ref{eqn:dmatrix}) and (\ref{eqn:Green})
are solved self-consistently. Then, the
total energy is given by
$$
E=
\sum_n
\frac{1}{\Omega_{\rm BZ}}
\int d {\bf k}
\tilde{\varepsilon}_{n{\bf k}} \Theta \left( \varepsilon_F - \tilde{\varepsilon}_{n{\bf k}} \right)
-\frac{1}{2} \sum_{\bm{\tau}} \sum_{ \mu_1 \mu_2 }
{\cal U}_{\bm{\tau}}^{\mu_1 \mu_2}
n_{\bm{\tau}}^{\mu_1 \mu_2}.
$$

  According to the constraint-LSDA calculations,\cite{PRB94} $J$ for Mo is of the order
of $0.5$ eV, and is not sensitive to details of the crystal environment
in solids. $B$ can be
estimated from $J$ using the ratio $F^4/F^2$$\simeq$$0.63$, which
holds for the Slater integrals in the atomic limit. This yields
$B$$\simeq$$0.06$ eV. The Coulomb $U$ is treated as the parameter
in order to consider different
scenarios, covering both metallic and insulating behavior of
$A_2$Mo$_2$O$_7$. The constraint-LSDA calculations for Mo compounds typically yield
$U$$\approx$$3.0$ eV.\cite{PRB94} This value can be further reduced by allowing for
the (proper) $e_g$ electrons to participate in the screening of the on-site Coulomb
interactions between $t_{2g}$ electrons.\cite{PRB96}

  For the practical calculations along this line, it is convenient to use a short
atomic-like basis set. For these purposes, $\varepsilon_{n{\bf k}}$ and
$|n {\bf k}\rangle$ have been first calculated using the nearly-orthogonal version
of the linearized-muffin-tin orbital method in the atomic-spheres approximation
(ASA-LMTO).\cite{LMTO}
The atomic orbitals $|{\bf R} \bm{\tau} \mu \rangle$ have been identified with
the basis orbitals of the LMTO method.
Atomic spheres radii have been chosen so to reproduce the electronic structure obtained
by more accurate
FLAPW method,\cite{Hamada,Lars,Singh1} especially for $t_{2g}$
bands located near the Fermi level. We have also added
a number of empty spheres at the $8a$ and $32e$ positions of the $Fd\overline{3}m$
lattice. All calculations have been performed on the mesh of
12$\times$12$\times$12 ${\bf k}$-points in the first Brillouin zone.

  Formally, our approach is similar to the rotationally-invariant
LDA$+$$U$,\cite{PRL98}
except that we apply it only to a limited number of bands, directly
responsible for the properties of $A_2$Mo$_2$O$_7$ compounds.
This will certainly simplify the calculations.
In addition,
we get rid of several artifacts of the conventional LDA$+$$U$
method, such as the substraction of the double-counting terms, which is
a very ambiguous procedure by itself, and in the number of cases leads to
a systematic error.\cite{PRB98}

\begin{center}
\section{\label{sec:DOS} Changes of the Electronic Structure
Caused by the on-site
Coulomb Interactions}
\end{center}

\begin{figure}
\centering \noindent
\resizebox{10cm}{!}{\includegraphics{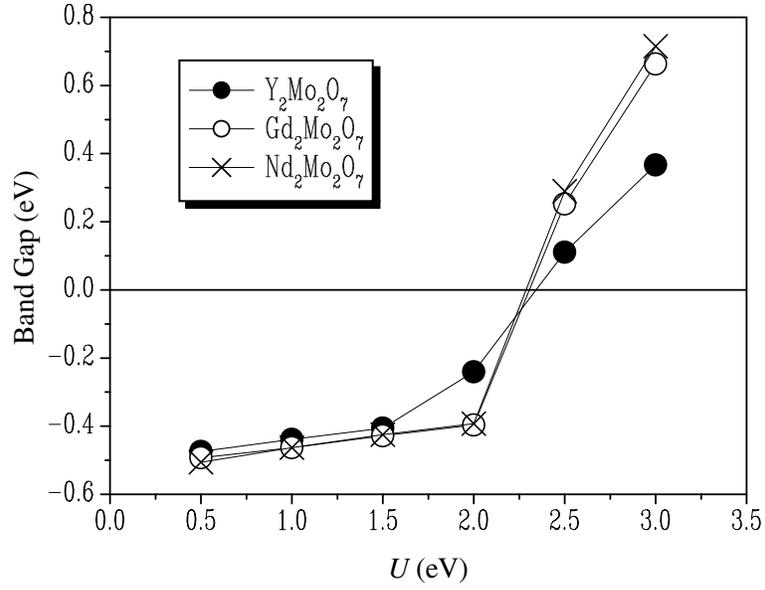}}
\caption{The band gap
as a function of Coulomb $U$. Here, the band gap is defined as the distance
between the bottom of 9th and the top 8th band, so that for
$U$$\leq$$2.0$ eV there is an overlap between these two bands, corresponding
to the negative value of the band gap, while
$U$$\geq$$2.5$ eV opens the real gap.}
\label{fig.gap}
\end{figure}

  Let us first discuss results without SOI, assuming FM ordering of the Mo spins.
According to the LSDA calculations (Fig.~\ref{fig.DOSt2g}), the majority
($\uparrow$) spin $a_{1g}$ band is fully occupied and the Fermi level crosses
the doubly-degenerate $e_g'$ band. Therefore, it is clear that at some point the
Coulomb $U$ will split the $e_g'$ band and form an insulating solution with
the spontaneously
broken $Fm\overline{3}d$ symmetry. Such situation occurs between $U$$=$ 2.0 and 2.5 eV
for all considered compounds (Fig.~\ref{fig.gap}, the AFM alignment of the Mo spins
may open the band gap even for smaller $U$, that well correlates with the fact that
all SG compounds are insulators).

\begin{figure}
\centering \noindent
\resizebox{7.5cm}{!}{\includegraphics{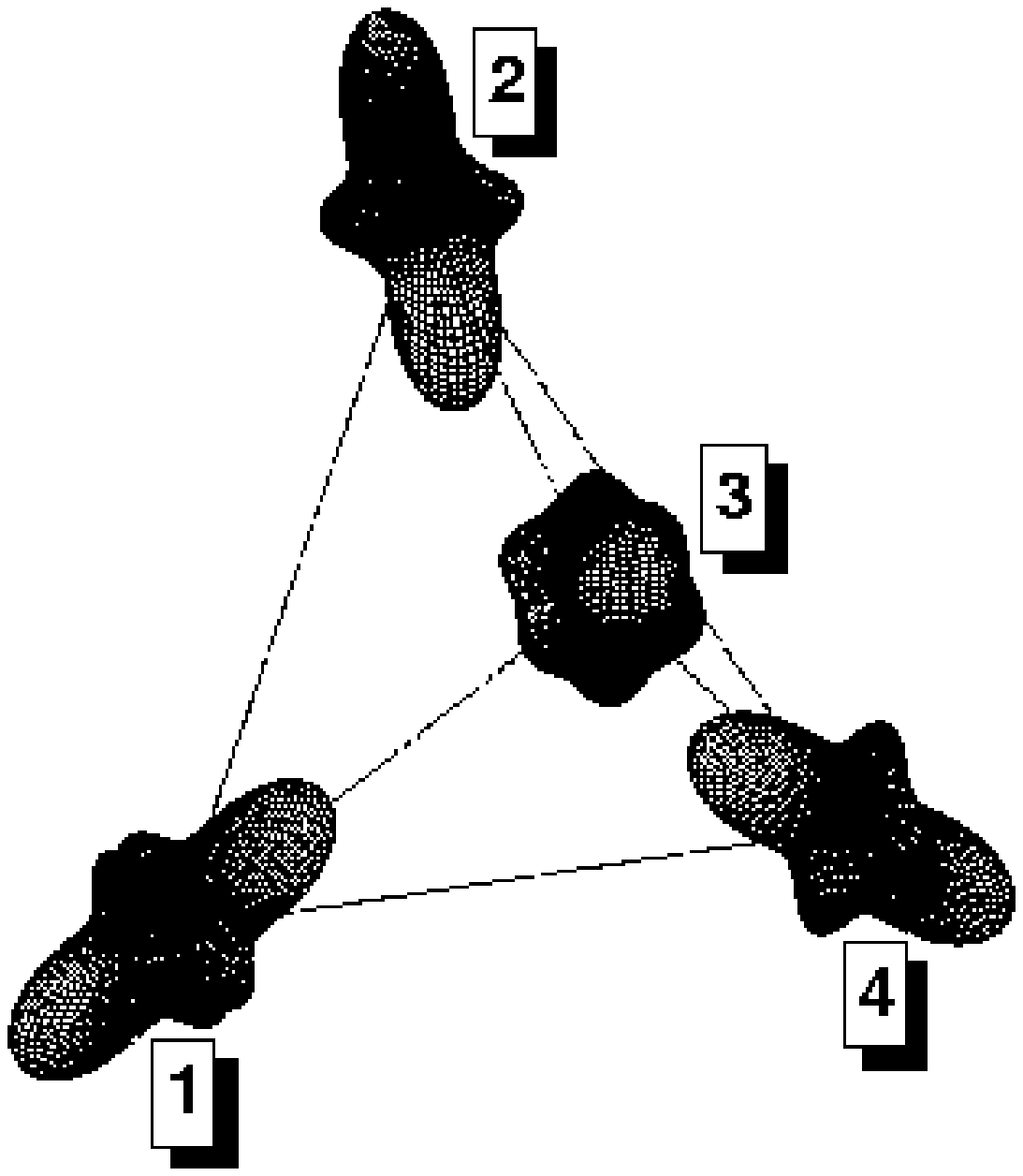}}
\caption{Orbital ordering in the ferromagnetic phase of Y$_2$Mo$_2$O$_7$
obtained
for $U$$=$$0.5$ eV.}
\label{fig.OOYF05}
\end{figure}

  In the metallic regime realized for small $U$, the densities of states are
similar to those in LSDA (Figs.~\ref{fig.DOSLSDA} and \ref{fig.DOSt2g}),
and we will not show them here again. We only mention that the major effect
of $U$ is the shift of the $\uparrow$-spin $a_{1g}$ band to the low-energy part of the
spectrum relative to the $e_g'$ band.
Corresponding orbital ordering
(the distribution of Mo($4d$) electron densities)
is shown in Fig.~\ref{fig.OOYF05}.
It comes exclusively from the local trigonal distortions of the oxygen octahedra and
represents the alternating $a_{1g}$ orbital densities in the background of degenerate
$e_g'$ orbitals.

\begin{figure}
\centering \noindent
\resizebox{10cm}{!}{\includegraphics{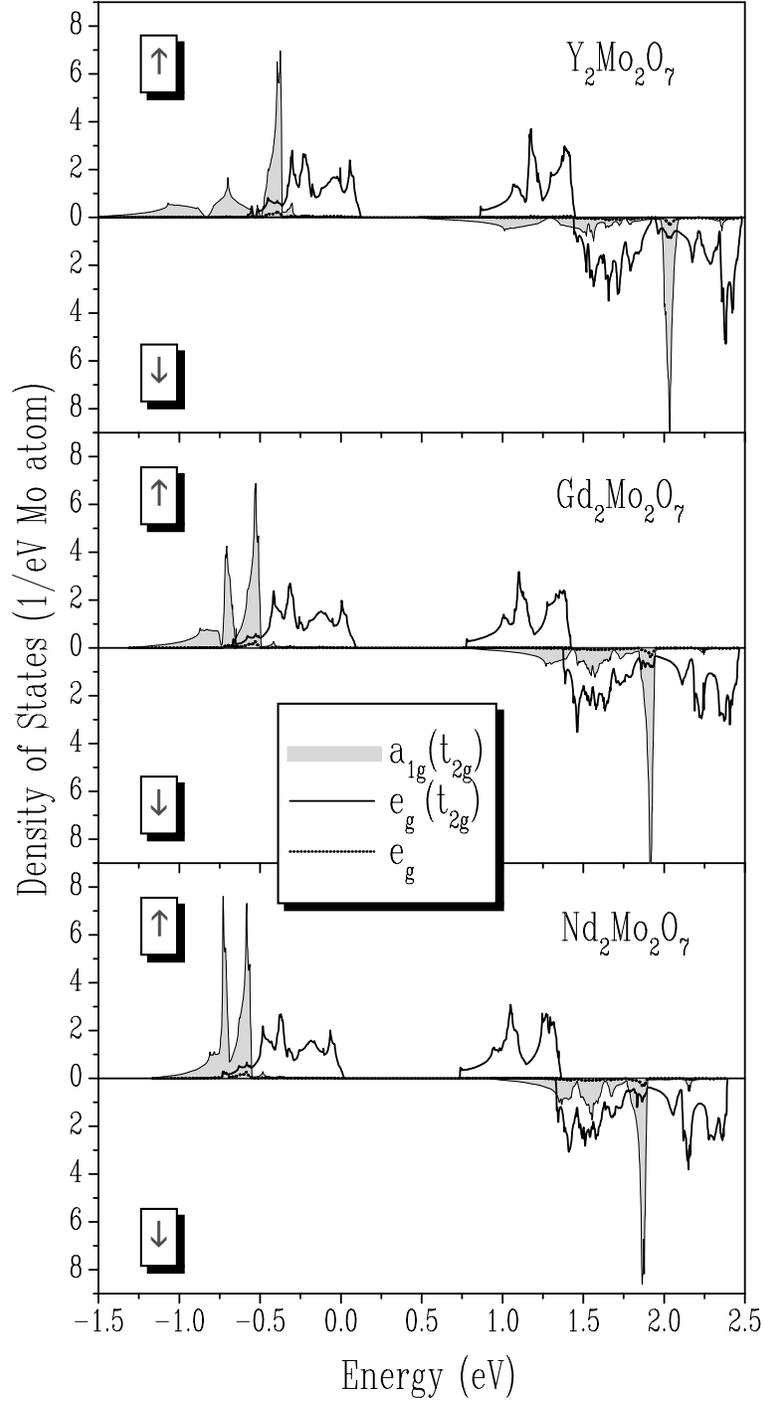}}
\caption{Distribution of the Mo($4d$) states
obtained in model Hartree-Fock
approach for $U=3.0$ eV.}
\label{fig.DOS30}
\end{figure}

\begin{figure}
\centering \noindent
\resizebox{7.5cm}{!}{\includegraphics{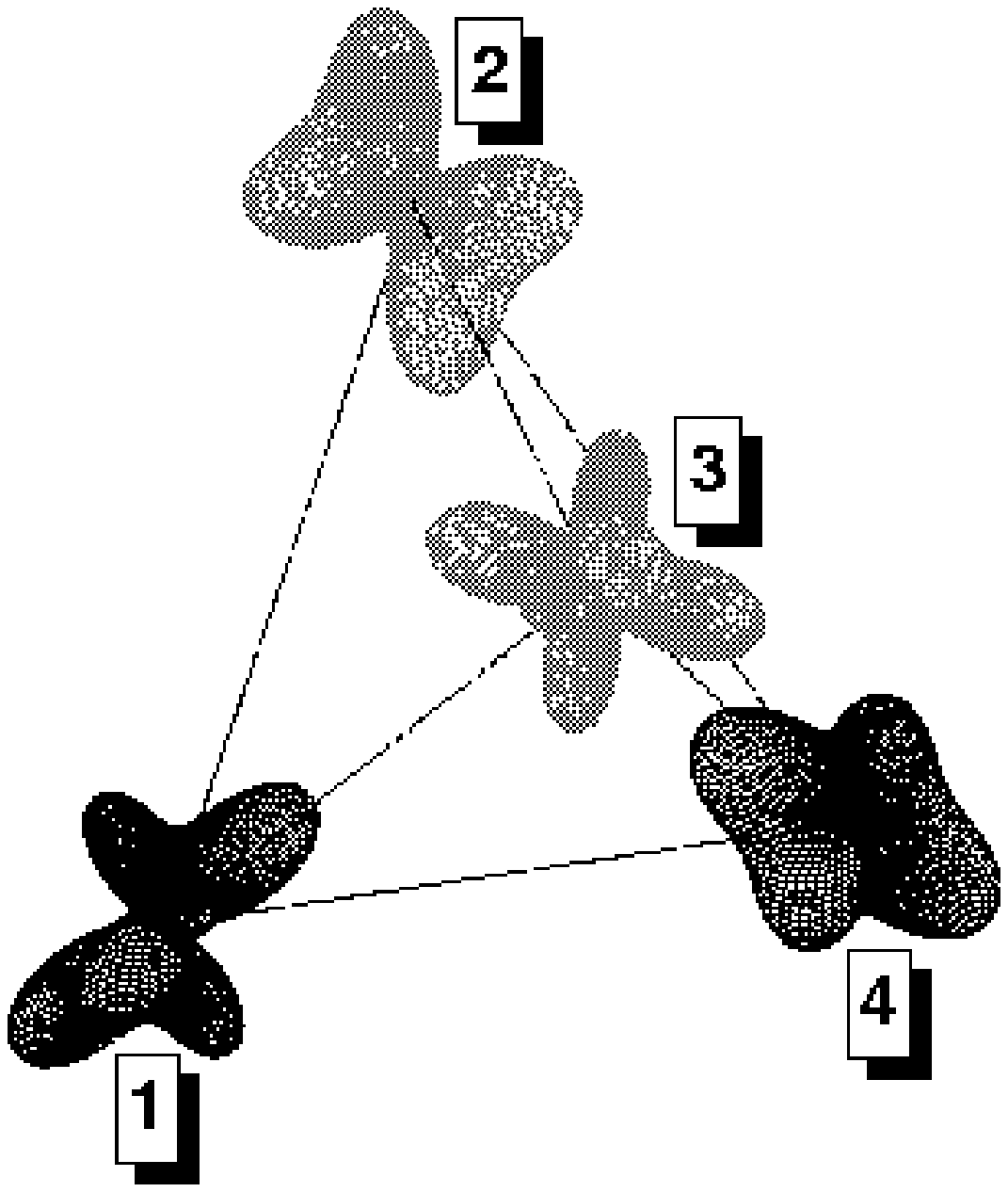}}
\caption{Orbital ordering in the ferromagnetic
phase of Y$_2$Mo$_2$O$_7$
obtained
for $U$$=$$3.0$ eV. Two orbital sublattices are shown by black and grey colors.}
\label{fig.OOYF30}
\end{figure}

\begin{figure}
\centering \noindent
\resizebox{7.5cm}{!}{\includegraphics{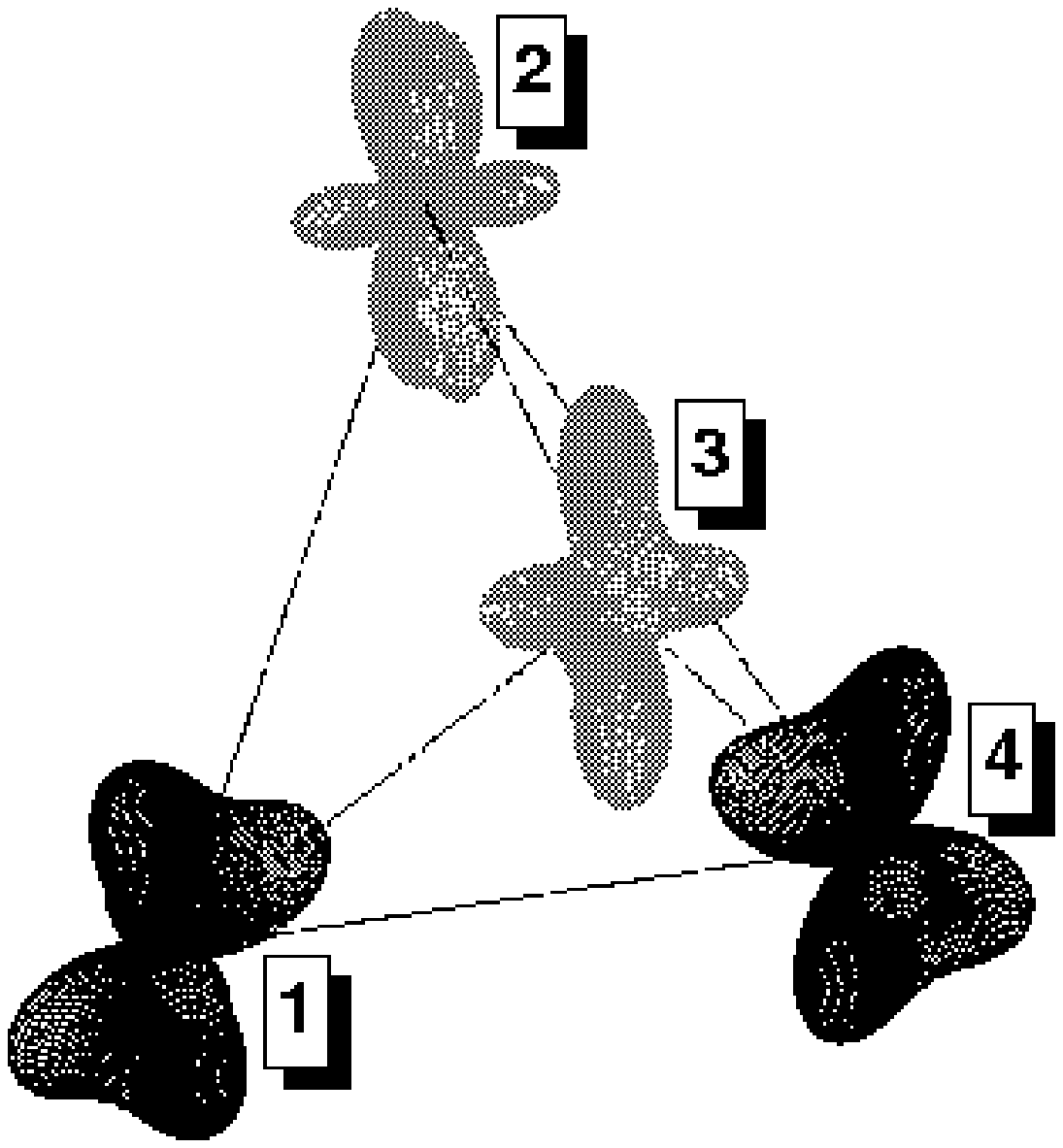}}
\caption{Orbital ordering
in the antiferromagnetic phase of Y$_2$Mo$_2$O$_7$
obtained
for $U$$=$$3.0$ eV. Two spin sublattices are shown by black and grey colors.}
\label{fig.OOYA30}
\end{figure}

  Typical densities of states in the insulating phase are shown in Fig.~\ref{fig.DOS30}.
The $a_{1g}$ band has characteristic for pyrochlores
three-peak structure.\cite{comment.4}
Smaller band gap in the case of Y$_2$Mo$_2$O$_7$ (Fig.~\ref{fig.gap}) is related with
broad minority ($\downarrow$) spin $a_{1g}$ band, whose tail spreads below the
unoccupied $\uparrow$-spin $e_g'$ band. The distribution of the $e_g'$ states is
very similar for all three compounds. Contrary to the metallic state, the
orbital ordering is determined not only by local trigonal distortions, but also
the form of the superexchange (SE) interactions between {\it nn} Mo sites, and minimizes
the energy of these interactions.\cite{KugelKhomskii}
Generally, it depends on the magnetic state. Two typical examples for the FM
and AFM (obtained after the flip of the
spin moments at the sites 2 and 3) phases
are shown in Figs.~\ref{fig.OOYF30} and
\ref{fig.OOYA30}, respectively. The building blocks of the AFM phase are the FM chains
of the atoms ...-1-4-1-4-.. and ...-2-3-2-3-..., which are coupled antiferromagnetically.
As expected for the FM spin ordering,\cite{KugelKhomskii}
the $e_g'$ orbitals tends to order ''antiferromagnetically'' and
form two Mo sublattices. Corresponding eigenfunctions of the occupied
$e_g'$ orbitals, obtained after the diagonalization of the density matrix
{\it in the local coordinate frame}, are given by
$|o_{1,4}\rangle$$\simeq$$\alpha|xy\rangle$$+$$\beta|yz\rangle$$+$$\gamma|zx\rangle$
for the sites 1 and 4, and
$|o_{2,3}\rangle$$\simeq$$\alpha|xy\rangle$$+$$\gamma|yz\rangle$$+$$\beta|zx\rangle$
for the sites 2 and 3, with the coefficients $\alpha$$\simeq$$0.600$,
$\beta$$\simeq$$0.193$, and $\gamma$$\simeq$$-$$0.769$. Clearly, this orbital
ordering breaks the $Fm\overline{3}d$ symmetry: if the sites belonging to the
same sublattice can still be transformed to each other using the symmetry operations
of the $Fm\overline{3}d$ group, the sites belonging to different sublattices -- cannot.
This will generally lead to the anisotropy of electronic properties, including the
anisotropy of {\it nn} magnetic interactions, which will be
discussed in the next section.

  The AFM spin ordering coexists with the FM orbital ordering for which
$\alpha$$\simeq$$-$$0.774$, and $\beta$$=$$\gamma$$\simeq$$0.442$
for all
Mo sites, in the local coordinate frame.\cite{comment.5}
It breaks the $Fm\overline{3}d$ symmetry in the spin sector, but not
in the orbitals one.

\ \

\section{\label{sec:exchange} Nearest-Neighbor Exchange Interactions}

  Parameters of exchange interactions between sites $\bm{\tau}$
and $\bm{\tau}'$ can be calculated as\cite{Liechtenstein}
\begin{equation}
J_{\bm{\tau} \bm{\tau}'} = \frac{1}{2\pi} {\rm Im}
\int_{-\infty}^{\varepsilon_F} d \varepsilon {\rm Tr}_L \left\{
\widehat{\cal G}^\uparrow_{\bm{\tau} \bm{\tau}'}(\varepsilon)
\widehat{\Delta}_{\bm{\tau}'}
\widehat{\cal G}^\downarrow_{\bm{\tau}' \bm{\tau}}(\varepsilon)
\widehat{\Delta}_{\bm{\tau}} \right\},
\label{eqn:exchange}
\end{equation}
where $\widehat{\cal G}^{\uparrow,\downarrow}_{\bm{\tau} \bm{\tau}'}(\varepsilon)$
is the block of matrix elements of the Green function
with the spin $\uparrow$ or $\downarrow$:
$$
\widehat{\cal G}^{\uparrow,\downarrow}_{\bm{\tau} \bm{\tau}'} =
\frac{1}{2} {\rm Tr}_S \left\{(\widehat{1} \pm \widehat{\sigma}_z)
\widehat{\cal G}_{\bm{\tau} \bm{\tau}'} \right\},
$$
and $\widehat{\Delta}_{\bm{\tau}}$ is the magnetic part of the HF
potential:
$$
\widehat{\Delta}_{\bm{\tau}} = {\rm Tr}_S
\{ \widehat{\sigma}_z \widehat{\cal U}_{\bm{\tau}} \}.
$$
${\rm Tr}_{S(L)}$ is the trace running over the spin
(orbital) indices.

\begin{figure}
\centering \noindent
\resizebox{10cm}{!}{\includegraphics{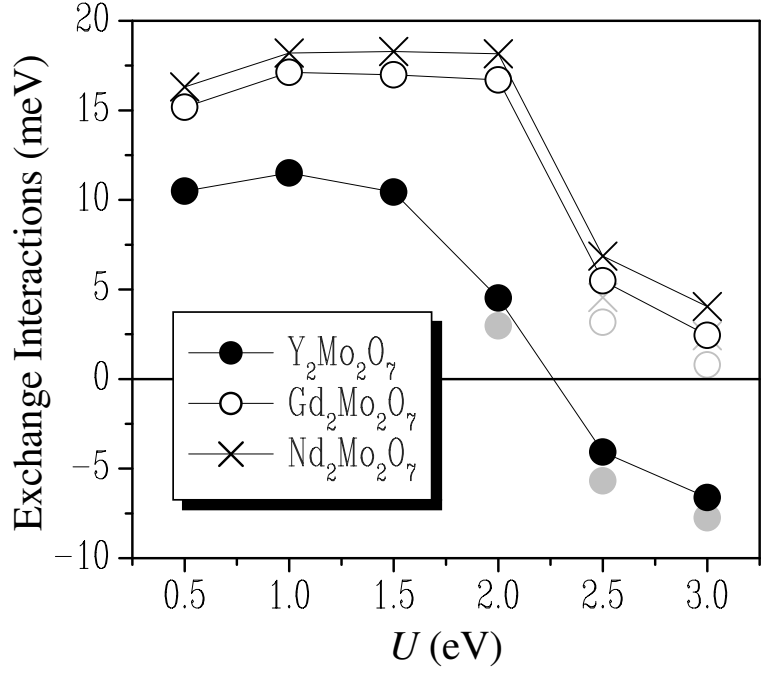}}
\caption{Nearest-neighbor exchange interactions calculated in the
ferromagnetic state. The orbital ordering realized for large
$U$ destroys the equivalence of some of the Mo-Mo bonds, that generally
leads to the inequality
$J_{12}$$=$$J_{13}$$\neq$$J_{14}$. Two such parameters, $J_{12}$
and $J_{14}$, are shown by dark and light symbols, respectively.}
\label{fig.exchange}
\end{figure}

\begin{figure}
\centering \noindent
\resizebox{10cm}{!}{\includegraphics{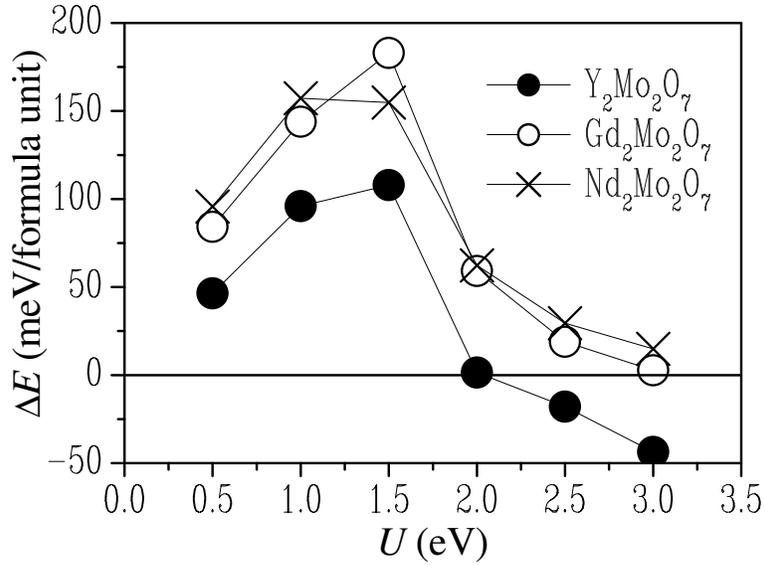}}
\caption{Total energy of the antiferromagnetic
state relative to the ferromagnetic state.}
\label{fig.etot}
\end{figure}

  Results for {\it nn} magnetic interactions for all considered compounds are
shown in Fig.~\ref{fig.exchange}, as a function of $U$.
We note the following.
\begin{enumerate}
\item
$J_{\bm{\tau} \bm{\tau}'}$, which are ferromagnetic for small $U$, exhibit a sharp
drop at the point of transition into the insulating state (see Fig.~\ref{fig.gap} for the
behavior of the band gap).
\item
There is a significant difference between Nd/Gd and Y based compounds: in the Y case,
the exchange parameters are almost rigidly shifted towards negative values, so that
the {\it nn} coupling become
antiferromagnetic in the insulating phase, while it
remains ferromagnetic in the case of Nd and Gd.
\end{enumerate}
Similar conclusions can be obtained from the analysis of total energies
shown in Fig.~\ref{fig.etot}.\cite{comment.2}

\begin{figure}
\centering \noindent
\resizebox{10cm}{!}{\includegraphics{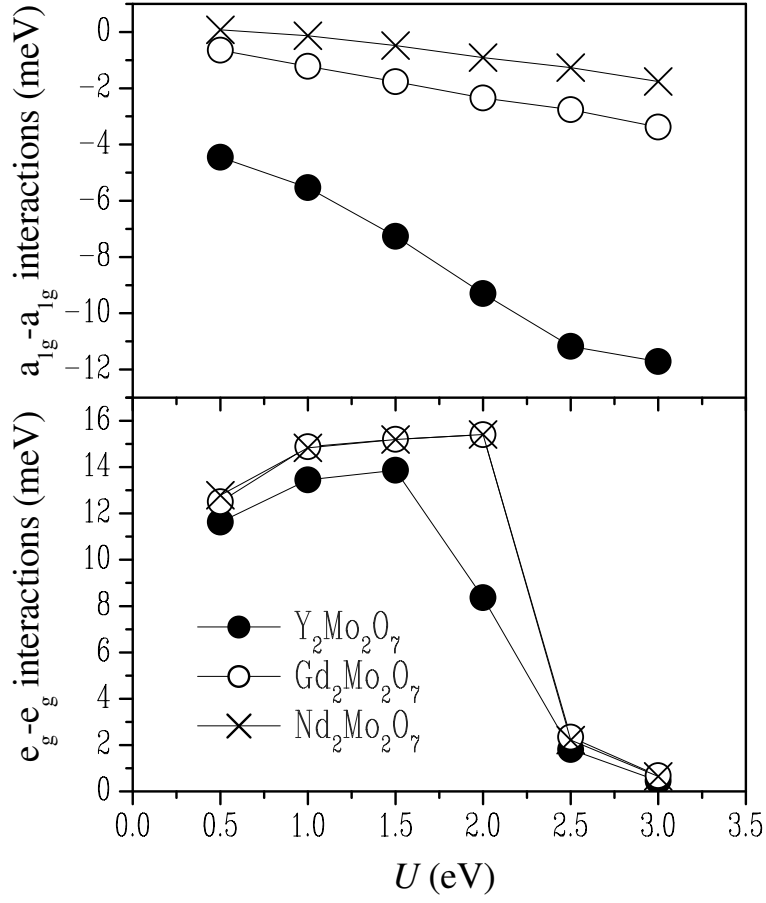}}
\caption{Contributions of $a_{1g}$ and $e_g'$
orbitals (coming from the $t_{2g}$ manifold in the local
coordinate frame) to the exchange constant $J_{14}$.}
\label{fig.jpartial}
\end{figure}

  The behavior can be easily understood by considering partial $a_{1g}$ and
$e_g'$ contributions to the {\it nn} exchange coupling, calculated after
transformation to the local coordinate frame at each site of the system
(Fig.~\ref{fig.jpartial}). Then, the main interactions are $a_{1g}$-$a_{1g}$
and $e_g'$-$e_g'$. The $a_{1g}$-$e_g'$ interaction is small and can be neglected.
The interaction between $t_{2g}$ and proper $e_g$ orbitals is of the order of
1.5-1.8 meV and only weakly depends on $U$ and the type of the
material. Therefore, it can be regarded as the weak FM background.

  Large FM $e_g'$-$e_g'$ interaction in the metallic regime is related with
the double exchange (DE) mechanism, which is the measure of the kinetic
energy for the itinerant $\uparrow$-spin $e_g'$ electrons
(note that for small $U$ all compounds are either
half-metallic or nearly half-metallic -- Fig.~\ref{fig.DOSLSDA}).\cite{Springer02}
Some increase of $J_{\bm{\tau} \bm{\tau}'}$ for small $U$ is caused by
the reduction of AFM SE contributions, which are proportional
to $1/U$. As long as the system is metallic, the DE interactions are not
sensitive to the value of $U$, and the FM coupling dominates.

  The transition into the insulating state is caused by the localization of the
$e_g'$ electrons on the atomic orbitals. This will reduce the kinetic energy
and suppress
the DE interactions, that explain the sharp drop of $J_{\bm{\tau} \bm{\tau}'}$
(Fig.~\ref{fig.exchange}).

  The main difference between Y and Nd/Gd based compounds is related with the
$a_{1g}$-$a_{1g}$ interactions. Since the $\uparrow$-spin $a_{1g}$ band is
fully occupied and the $\downarrow$-spin band is empty, the interactions are
antiferromagnetic and the underlying microscopic mechanism is the superexchange.
Since the SE coupling is proportional to the square of the $a_{1g}$ bandwidth,
this interaction is the largest in the case of Y, that explains the AFM
character of the total coupling realized in this compound for large $U$.\cite{comment.1}

  The anisotropy of the {\it nn} interactions, $\Delta J$$=$$J_{12}$$-$$J_{14}$,
caused by the orbital ordering in the FM state is of the order of 1-2 meV
(Fig.~\ref{fig.exchange}). This is a small value in comparison with
the sharp drop of $J_{\bm{\tau} \bm{\tau}'}$ caused by the Coulomb
interaction $U$. However, it can play an important role in the insulating regime,
where $\Delta J$ becomes comparable with
$J_{\bm{\tau} \bm{\tau}'}$.
For example, one interesting question is whether this $\Delta J$, of a purely
Coulombic origin, is sufficient to explain the appearance of the SG state.
In addition, $\Delta J$ has the same order of magnitude as the
single-ion anisotropy energy
(Sec.~\ref{sec:NC}) and the Mo-Gd exchange
interaction (Ref.~\onlinecite{MoGd_exchange}). Therefore, in the insulating state
there is no unique interaction which would control the magnetic behavior of the
$A_2$Mo$_2$O$_7$ compounds. All interactions are equally important, that represents
the main complexity on the way of theoretical description of these materials.

  The AFM spin ordering changes the orbital ordering in the direction which further
stabilizes the {\it nn} magnetic interactions,\cite{KugelKhomskii} and in a number
of cases may lead to the charge disproportionation within the FM chains.\cite{comment.7}
For example, $U$$=$$2.0$ eV appears to be sufficient to open the band gap ($\sim$$0.3$ eV)
in the AFM phase of Y$_2$Mo$_2$O$_7$. The solution is accompanied by the small charge
disproportionation, and the values of the spin magnetic moments at the four Mo sites
are $\mu_1^S$$=$$-$$\mu_3^S$$=$$1.4$ $\mu_B$ and $-$$\mu_2^S$$=$$\mu_4^S$$=$$1.5$ $\mu_B$.
Corresponding values of the inter-atomic exchange interactions are
$J_{12}$$=$$-$$0.8$, $J_{13}$$=$$-$$2.9$ and $J_{14}$$=$$2.6$ meV, i.e. the interactions
are antiferromagnetic in the bonds 1-2 and 1-3, and ferromagnetic in the bond 1-4.
Therefore, the obtained solution is locally stable.
The difference between $J_{12}$ and $J_{13}$ is caused by the charge disproportionation.
Further increase of $U$ stabilizes
the AFM coupling in all Mo-Mo bonds, giving rise to the geometrical
frustrations. Typical values of the exchange parameters obtained in the HF solution
for $U$$=$$3.0$ eV,
which does not show any sign of the charge disproportionation,
are $J_{12}$$=$$J_{13}$$=$$-$$10.7$ and $J_{12}$$=$$-$$2.4$ meV.

\section{\label{sec:NC} Non-collinear Magnetic Ordering}

  In this and subsequent sections we will discuss the effects of SOI on the electronic
and magnetic structure of the Mo pyrochlores. The matrix element of $\xi({\bf r})$
calculated in the basis of Mo($4d$) orbitals is of the order of
$\xi_{4d}$$=$$120$ meV. It defines the characteristic
energy gain after adding the SOI into the model HF Hamiltonian. The equilibrium
magnetic ordering depends on the competition between the isotropic exchange interactions,
considered in the previous section, and the single-ion anisotropies energy.
The isotropic exchange interactions tend to align Mo spins parallel to each other
(at least in the case of the FM coupling). However, due to
the peculiarities of the pyrochlore structure,
there is no unique direction of the spin magnetization which would satisfy the
single-ion anisotropy energies simultaneously for all Mo sites.
These anisotropies lead to the distortion of the collinear FM ordering.\cite{Moriya}

\begin{figure}
\centering \noindent
\resizebox{10cm}{!}{\includegraphics{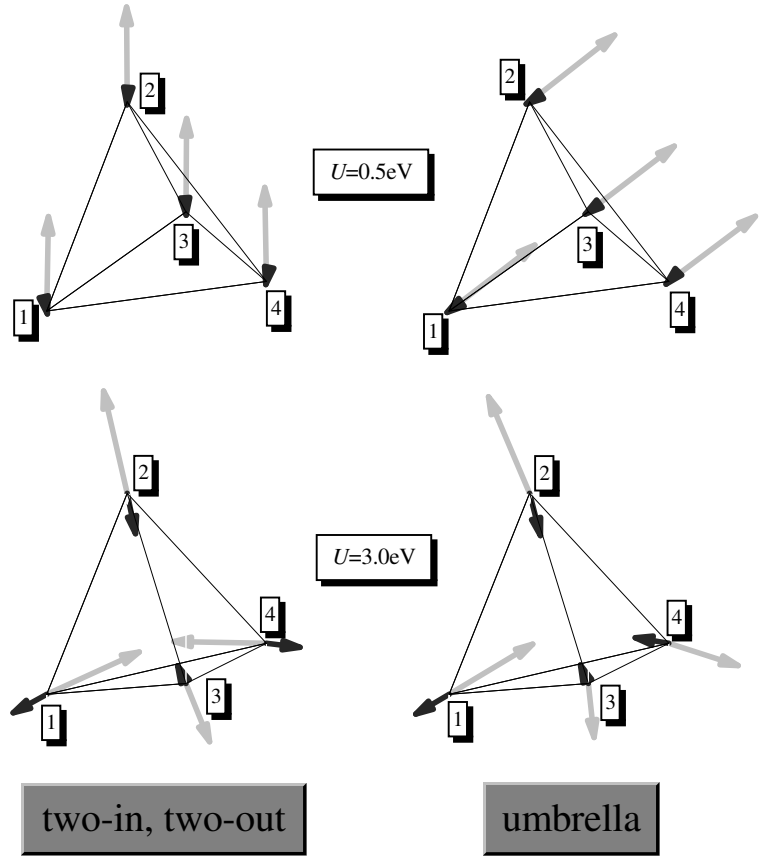}}
\caption{Two magnetic structures realized in the case of the ferromagnetic
coupling between nearest-neighbor Mo spins: ''two-in, two-out'' (left)
and ''umbrella'' (right). The relative orientation and the magnitude
of the spin (light vectors) and orbital (dark vectors) magnetic moments
corresponds to Gd$_2$Mo$_2$O$_7$ for two different values of the Coulomb
interaction $U$.}
\label{fig.fstructures}
\end{figure}

  In the case of FM coupling between {\it nn} spins, there are two magnetic structures,
which are typically discussed in the literature and can be obtained by distorting the
collinear
FM ordering along either $(0,0,1)$ or $(1,1,1)$ direction
(see Fig.~\ref{fig.fstructures}).
\begin{enumerate}
\item
The ''two-in, two-out'' structure, where the directions of the magnetic moments at the
four
Mo sites are given by: ${\bf e}_1$$=$$\frac{1}{\sqrt{1+\eta^4}}(\eta,\eta,1$$-$$\eta^2)$,
${\bf e}_2$$=$$\frac{1}{\sqrt{1+\eta^4}}($$-$$\eta,\eta,1$$-$$\eta^2)$,
${\bf e}_3$$=$$\frac{1}{\sqrt{1+\eta^4}}(\eta,$$-$$\eta,1$$-$$\eta^2)$, and
${\bf e}_4$$=$$\frac{1}{\sqrt{1+\eta^4}}($$-$$\eta,$$-$$\eta,1$$-$$\eta^2)$,
in terms
of the distortion $\eta$ from the perfect FM alignments.
\item
The ''umbrella'' structure: ${\bf e}_1$$=$$\frac{1}{\sqrt{3}}(1,1,1)$,
${\bf e}_2$$=$$\frac{1}{\sqrt{3(1+2\eta^2)}}(1$$-$$2\eta,1$$+$$\eta,1$$+$$\eta)$,
${\bf e}_3$$=$$\frac{1}{\sqrt{3(1+2\eta^2)}}(1$$+$$\eta,1$$-$$2\eta,1$$+$$\eta)$, and
${\bf e}_2$$=$$\frac{1}{\sqrt{3(1+2\eta^2)}}(1$$+$$\eta,1$$+$$\eta,1$$-$$2\eta)$.
\end{enumerate}

  Since the Mo($4d$) states are strongly polarized by the on-site Coulomb interactions,
the main contribution to the single-ion anisotropy energy comes from the
$\uparrow$-spin electrons. In this case there is an one-by-one correspondence
between the single-ion anisotropy energy and
the magnitude of the orbital magnetization, so that the direction which maximizes the
magnetization automatically minimizes the magnetic anisotropy energy.\cite{Bruno}
For the $t_{2g}$ electrons affected by the trigonal distortion,
the spontaneous orbital magnetization is determined by
the following factors:\cite{JMMM1998}
\begin{enumerate}
\item
The interaction between occupied $a_{1g}$ and unoccupied $e_g'$ states favors
the in-plane geometry of the orbital magnetization (i.e., perpendicular to the
trigonal axes). The latter is proportional to the ratio $\xi_{4d}/ \delta \varepsilon_{a_{1g}}$,
where $\delta \varepsilon_{a_{1g}}$ is the distance between
the position of the $a_{1g}$ states and
the unoccupied part of the spectrum. Clearly, this interaction will vanish
when the Coulomb $U$ is large.
\item
On the contrary, the interaction between $e_g'$ orbitals favors the easy axes
geometry. In the pure ionic limit, such solution is the subject for the degenerate
perturbation theory. Therefore, the wave functions and the
orbital magnetization do not depend on
$\xi_{4d}$. The SOI in such situation only
lifts the orbital degeneracy and
defines the symmetry of the $e_g'$ orbitals which will be further split
by $U$. The solution is accompanied by the large
orbital moment of the order of $1$ $\mu_B$, which is
parallel to the trigonal axis.
\end{enumerate}

  Thus, when $U$ is sufficiently large, the second mechanism will dominate and one can
naturally expect the easy axes scenario for the single-ion anisotropies. In the combination
with the FM coupling between {\it nn} spins, it yields the ''umbrella''
ground state.\cite{comment.6}

\begin{figure}
\centering \noindent
\resizebox{9cm}{!}{\includegraphics{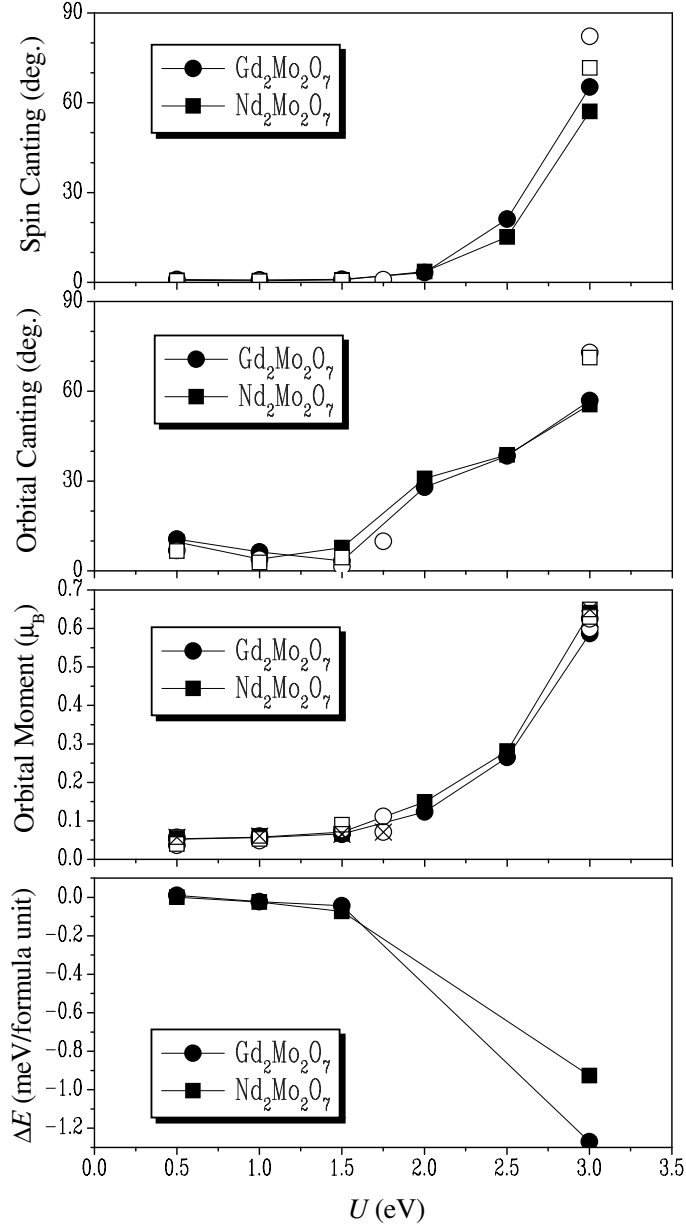}}
\caption{The angles formed by the
spin and orbital magnetic moments and either $(0,0,1)$ (''two-in, two-out'')
or $(1,1,1)$ (''umbrella'') axes; the absolute value of the
orbital moment; and the total energy of the ''umbrella'' structure relative to the
''two-in, two-out'' structure as a function
of Coulomb $U$. Filled and open symbols correspond to the ''two-in, two-out''
and ''umbrella'' structure, respectively.
In the case of ''umbrella'' structure, there are two different magnetic moments:
the ones at the site 1 are shown by regular symbols, $\bigcirc$ and $\Box$,
and the ones at the site 2, 3 and 4
(forming the top of the ''umbrella'') are shown by crossed symbols.
The spin magnetic moment is of the order of $1.5$-$1.6\mu_b$, and weakly
depends on $U$.}
\label{fig.canting}
\end{figure}

  The qualitative analysis is supported by results of direct calculations shown in
Fig.~\ref{fig.canting}. In the metallic regime, the isotropic double exchange
interaction between Mo spins clearly dominates over the single-ion
anisotropy energies and the canting of spin magnetic moments is small
(both for the ''two-in, two-out'' and ''umbrella'' structure). The orbital moments
are not particularly strong ($<$$0.1$ $\mu_B$ per Mo site). They are canted off
the global FM direction by about $10^\circ$.

  In the insulating state, the isotropic exchange interactions are
suppressed and become comparable with the single-ion anisotropy energies,
which are enhanced. This
results in the large canting of the magnetic moments, which can be very different
for the spin and orbital counterparts (in the other words, the spin and orbital
moments are not collinear with respect to each other).\cite{Sandratskii,PRB97}
The reason for this difference is the following. Directions of the orbital magnetic
moments are determined mainly by the single-ion anisotropy terms, whereas the
interactions between moments located at different sites
are typically much smaller. Therefore, the orbital moments
always tend to align first along the crystallographic directions (the trigonal axes,
in our case). On the contrary, alignment of the spin magnetic moments is strongly
affected by the inter-site interactions, which generally lead to a smaller
canting.

  For large $U$ ($\sim$3 eV), the single-ion anisotropies start to
dominate over the isotropic exchange interactions. In this limit, the orbital magnetic
moments are almost perfectly aligned along the trigonal axes
(that corresponds to the 55$^\circ$ and 70$^\circ$
canting for the ''two-in, two-out'' and ''umbrella'' structure, respectively).
The value of the local orbital magnetic
moment is about 0.6 $\mu_B$, which is smaller than 1 $\mu_B$
expected in the pure atomic limit.

  The ''umbrella'' structure has lower energy
(perhaps except the region of very small $U$). The energy difference between the
''two-in, two-out'' and ''umbrella'' structures is small in the metallic
regime ($\sim$$0.05$ meV per formula unit, which corresponds to the coercive
field of 1 Tesla). However,
it becomes large in the insulating state, where the interactions between
the $e_g'$ orbitals clearly dominate.

\begin{figure}
\centering \noindent
\resizebox{9cm}{!}{\includegraphics{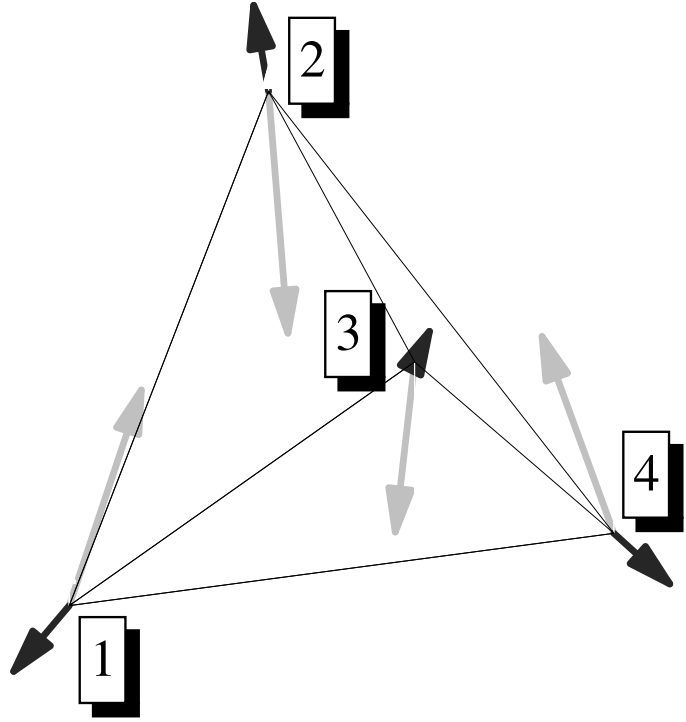}}
\caption{Magnetic structure in the case of antiferromagnetic
coupling between nearest-neighbor Mo spins.
The relative orientation and the magnitude
of the spin (light vectors) and orbital (dark vectors) magnetic moments
corresponds to Y$_2$Mo$_2$O$_7$ for $U$$=$$3.0$ eV.}
\label{fig.astructure}
\end{figure}

  In the case of AFM coupling between {\it nn} spins, the single-ion anisotropies
lift, at least partially, the high degeneracy of the magnetic ground state in
the frustrated pyrochlore lattice. The equilibrium magnetic
structure realized in model HF calculations for Y$_2$Mo$_2$O$_7$ can be obtained by the
inversion of the magnetic moments at the sites 2 and 3 of the ''two-in, two-out'' structure
(Fig.~\ref{fig.astructure}). For $U$$=$$3.0$ eV, the values of spin and orbital
magnetic moments at the Mo sites are $1.44$ and $0.51$ $\mu_B$, respectively. They are canted off the
cubic (0,0,1) axis by correspondingly $17^\circ$ and $40^\circ$. The AFM interactions
are compromised with the tetrahedral geometry of the pyrochlore lattice by forming the
AFM coupling between two of the three projections of the {\it nn} spins in
each of the Mo-Mo bonds, while the remaining projection is coupled ferromagnetically.

\begin{center}
\section{\label{sec:Hall} Anomalous Hall Conductivity and Magneto-Optical Effect}
\end{center}

  The conductivity tensor,
$\widehat{\sigma}(\omega)$$=$$\widehat{\sigma}^{\rm reg}(\omega)$$+$$\widehat{\sigma}^{\rm D}(\omega)$,
can be calculated
from the eigenvalues, $\tilde{\varepsilon}_{{\bf k}n}$, and the eigenfunctions,
$|\widetilde{{\bf k}n} \rangle$,
of the HF Hamiltonian using the Kubo formula,\cite{method_optica,method_moptica} where
the inter-band (regular) and intra-band (Drude) contributions
are given by (in Ry units)
\begin{equation}
\sigma_{\alpha \beta}^{\rm reg}(\omega)=\frac{4i}{V} \sum_{\bf k} \sum_{nn'}
\frac{f_{{\bf k}n'}-f_{{\bf k}n'}}{\omega_{n'n}}
\left[
\frac{\pi^\alpha_{nn'} \pi^\beta_{n'n}}{\omega-\omega_{n'n}+i/\tau} +
\frac{(\pi^\alpha_{nn'} \pi^\beta_{n'n})^\ast}{\omega+\omega_{n'n}+i/\tau} \right]
\label{eqn:sinter}
\end{equation}
and
\begin{equation}
\sigma_{\alpha \beta}^{\rm D}(\omega)=\frac{8i}{V} \frac{1}{\omega+i/\tau} \sum_{\bf k} \sum_n
 \pi^\alpha_{nn} \pi^\beta_{nn} \delta(\varepsilon_F-\tilde{\varepsilon}_{{\bf k}n}),
\label{eqn:sintra}
\end{equation}
respectively. In these equations,
$f_{{\bf k}n}$ is the zero-temperature Fermi distribution function,
$\pi^\alpha_{nn'}$$=$$\langle \widetilde{{\bf k}n} | (-i\nabla_\alpha) | \widetilde{{\bf k}n'} \rangle$
is the matrix element of momentum operator, $\omega_{n'n}$$=$$\tilde{\varepsilon}_{{\bf k}n'}$$-$$
\tilde{\varepsilon}_{{\bf k}n}$, $V$ is the unit cell volume, and $\tau$ is the phenomenological
relaxation time ($1/\tau$$\simeq$$68$ meV) introduces in order
to perform summation over the discrete mesh of ${\bf k}$-points.
The diagonal elements of $\widehat{\sigma}^{\rm D}(\omega)$ can be also expressed through the
plasma frequencies, $\omega^\alpha_p$,
as $\sigma^{\rm D}_{\alpha \alpha}$$=$$\frac{(\omega^\alpha_p)^2}{4\pi}
\frac{i}{\omega+i/\tau}$.

  The conductivity tensor has the form
\begin{equation}
\widehat{\sigma} =
\left(
\begin{array}{ccc}
\sigma_{xx}    &  \sigma_{xy}  & 0            \\
-\sigma_{xy}   &  \sigma_{xx}  & 0            \\
    0          &    0          & \sigma_{zz}  \\
\end{array}
\right),
\label{eqn:tensor}
\end{equation}
both for the ''two-in, two-out'' and ''umbrella'' structure, if the direction of the
$z$ axis is chosen as $(0,0,1)$ and $(1,1,1)$, respectively.

\begin{figure}
\centering \noindent
\resizebox{9cm}{!}{\includegraphics{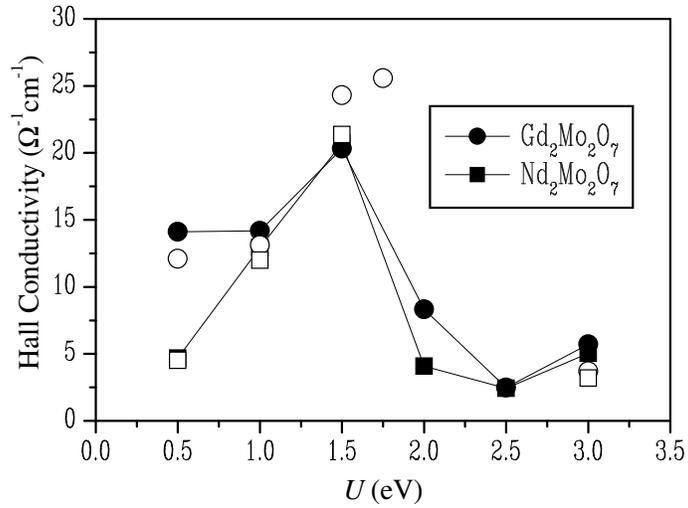}}
\caption{Hall Conductivity in Gd$_2$Mo$_2$O$_7$ and Nd$_2$Mo$_2$O$_7$
as a function of Coulomb $U$.
Filled and open symbols correspond to the ''two-in, two-out''
and ''umbrella'' structure, respectively.}
\label{fig.Hall}
\end{figure}

  The Hall conductivity, defined as $\sigma_H$$=$${\rm Re} \sigma_{xy}(0)$,
is shown in Fig.~\ref{fig.Hall} as a function of $U$. As expected,\cite{comment.14}
the effect is caused
by the inter-band transitions, while all intra-band contributions vanish, even in
the metallic state.
$\sigma_H$ takes the maximal
value around $U$$=$$1.5$ eV, which compromises between the value of the
orbital magnetization
and the metallic behavior of the system. On the one hand, $U$ is large enough
to unquench the orbital magnetic moment of the order of 0.08 $\mu_B$
per Mo site. On the other hand it is
small enough in order
to keep the system in the metallic regime (see Fig.~\ref{fig.gap}). The
factor-two increase of $\sigma_H$ from $U$$=$$0.5$ to $1.5$ eV well correlates with
similar increase of the orbital magnetic moment (Fig.~\ref{fig.canting}),
suggesting that both effects have relativistic origin.
The behavior coexists with nearly collinear FM ordering
realized in the metallic regime.
The transition into the insulating state, which is accompanied by the large canting of
the magnetic moments, results in the abrupt drop of $\sigma_H$.\cite{comment.8}
Therefore, we conclude that the large $\sigma_H$ observed in the experiment is
due to the orbital magnetic moments induced by
the SOI in the nearly collinear FM state.\cite{KL}
The absolute value of $\sigma_H$, $\sim$$25$ $\Omega^{-1}{\rm cm}^{-1}$, obtained in the
calculations is in a good agreement with the experimental data for
Nd$_2$Mo$_2$O$_7$.\cite{Taguchi3} However, we would like to emphasize again that the
SOI alone is not sufficient to obtain such a large values of $\sigma_H$, and
it is essential to consider the on-site Coulomb interactions beyond LSDA, which
play a very important role in the problem and enhance the effect of the
SOI.\cite{PRL98}

  According to our calculations, $\sigma_{xy}$, in the geometry which
specify the form of the tensor (\ref{eqn:tensor}), only weakly depends on the magnetic
structure and the direction of the net magnetization. Large anisotropy of $\sigma_H$
observed in the experiment\cite{Taguchi3,Yoshii} is presumably related with the details
of the experimental set-up and reflects the ''$\frac{1}{\sqrt{3}}$-law''.
For example, if the electric filed is directed along the $(1,0,0)$ axis, and the
current is measures along the $(0,1,0)$ axis, the effect is proportional to
$\sigma_{xy}/\sqrt{3}$ in the case of the umbrella structure, that roughly
corresponds to the experimental situation. Similar arguments apply for the
magneto-optical Kerr rotation which will be discussed below.

  If our scenario is correct, and large AHE in the Mo pyrochlores
can be explained by the inter-band transitions in the presence of
the SOI (i.e., by the conventional means), it is natural to
expect similar enhancement for other phenomena related with the SOI. Below we
consider the complex Kerr effect, which can be also regarded as a possible check
for our theory in the future.

  The Kerr rotation ($\theta_K$) and Kerr ellipticity ($\epsilon_K$) can be
found from the expression,\cite{Uspenskii}
\begin{equation}
\frac{1 -\tan \epsilon_K}{1 +\tan \epsilon_K} e^{2i\theta_K} =
\frac{1-n_+}{1+n_+} \frac{1+n_-}{1-n_-},
\label{eqn:Kerr}
\end{equation}
which is formally exact and valid for arbitrary $\theta_K$ and $\epsilon_K$.
$n_\pm$$=$$\sqrt{\varepsilon_{xx} \pm i \varepsilon_{xy}}$ are the complex
refractive indices, in terms of matrix elements of the dielectric tensor,
$\varepsilon_{\alpha \beta}(\omega)$$=$$\delta_{\alpha \beta}$$+$$\frac{4\pi i}{\omega}
\sigma_{\alpha \beta}(\omega)$.

\begin{figure}
\centering \noindent
\resizebox{9cm}{!}{\includegraphics{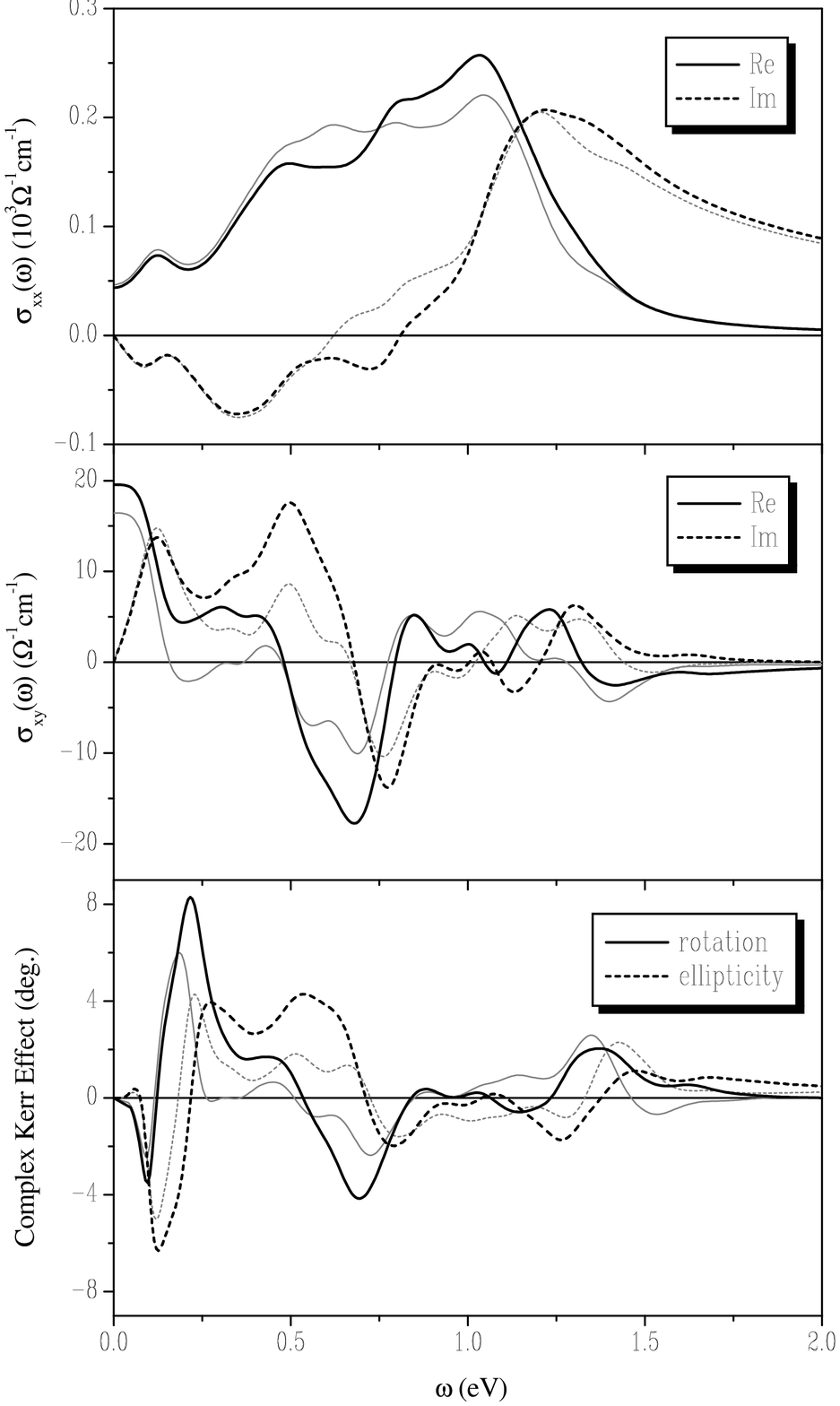}}
\caption{Matrix elements of optical conductivity and the Complex
Kerr effect for Gd$_2$Mo$_2$O$_7$ corresponding to the ''umbrella''
ground state realized for $U$$=$$1.5$ eV. Black lines show total spectra.
Gray lines are
obtained after excluding the $a_{1g}$ orbitals from the wave functions
(and roughly correspond to the partial contributions of the $e_g'$
bands).}

\label{fig.conductivity}
\end{figure}

  Results of these calculations for Gd$_2$Mo$_2$O$_7$ are shown in Fig.~\ref{fig.conductivity},
together with the $\omega$-dependence of the matrix elements of the optical conductivity.
We consider the ''umbrella'' structure
obtained for $U$$=$$1.5$ eV, which corresponds to the maximal
rotation. Calculations for Nd$_2$Mo$_2$O$_7$ and
the ''two-in, two-out'' structure
yield very similar results.
We also discuss partial contributions of the $a_{1g}$ and $e_g'$ states, and show
the spectra obtained after removing the $a_{1g}$ orbitals from the basis.\cite{comment.9}

  Apart from the HF model itself, we do not use any adjustable parameters. For example,
the intra-band Drude contribution to the optical conductivity has been obtained in
the calculations.

  The diagonal part of $\widehat{\sigma}(\omega)$ comes mainly from the inter-band
$e_g'$$\rightarrow$$e_g'$ transitions. The contribution of the $a_{1g}$ states is
rather small. The absorptive (Re) part of $\sigma_{xx}(\omega)$ has a maximum around 1 eV.
Both the peak position and its intensity roughly agree with the experimental data,\cite{Taguchi2}
tough the latter reveal an additional structure in the low-$\omega$ energy part of the spectrum,
which can be due to the quantum fluctuation effects,\cite{TokuraNagaosa}
beyond the mean-field approximation
considered in our work.\cite{comment.11}

  The contribution of the $a_{1g}$ states to
$\sigma_{xy}$ is very important, and can bring up to 50\% of the
spectral weight at certain values of $\omega$. However, the low-$\omega$ part of
$\sigma_{xy}$ comes mainly from the inter-band
$e_g'$$\rightarrow$$e_g'$ transitions,
which are responsible for almost 90\% of the Hall conductivity.
The large intensity of $\sigma_{xy}$ is due to the
combination of three factors: (i) large SOI at the Mo sites; (ii) the enhancement
of the SOI by the on-site Coulomb interactions, which is especially efficient for
the degenerate $e_g'$ states; (iii) the half-metallic electronic structure of
Gd$_2$Mo$_2$O$_7$ and Nd$_2$Mo$_2$O$_7$: since only $\uparrow$-spin states
contribute to the low-$\omega$ part of
$\sigma_{xy}$, the cancellation between
different spin channels does not take place.\cite{Irkhin}

  The Kerr rotation spectrum, calculated using Eg.~(\ref{eqn:Kerr}), has two
structures. The first one is close to the plasma edge ($\hbar \omega$$\simeq$$0.3$ eV),
and is directly related with the plasma enhancement effect.\cite{Uspenskii}
The exceptionally large value of $\theta_K$ ($\sim$$8^\circ$) is also related with
large
$\sigma_{xy}$ in this region, which correlates with the large value
of the Hall conductivity. The second structure ($\theta_K$$\sim$$-$$4^\circ$)
near $0.7$ eV is related with the form of ${\rm Re}\sigma_{xy}(\omega)$,
which has a deep in this region as a joint effect of both $e_g'$$\rightarrow$$e_g'$ and
$a_{1g}$$\rightarrow$$e_g'$ inter-band
transitions. Smaller $U$ would yield
somewhat smaller estimates for the rotation angles (correspondingly $4$ and $-$$1.5^\circ$ for
$U$$=$$0.5$ eV). Nevertheless, the FM pyrochlores $A_2$Mo$_2$O$_7$ seem to
be rather promising materials for the magneto-optical applications.

  The anomalous Hall and Kerr effects simultaneously
vanish in the AFM phase, due to the cancellation
of contributions coming from different Mo sublattices
in the macroscopic sample.\cite{Dzyaloshinskii}

\ \

\section{\label{sec:summary}Discussions and Summary}

  We have presented results of model HF calculations for pyrochlores $A_2$Mo$_2$O$_7$
($A$$=$ Y, Gd, and Nd). The model combines details of LDA electronic structure for the Mo($t_{2g}$)
bands located near the Fermi level,
the spin-orbit interaction, and the on-site Coulomb interactions amongst the Mo($4d$)
electrons.

  The main results can be summarized as follows.

  $\bullet$ The magnetic ground state of the $A_2$Mo$_2$O$_7$ compounds
depends on the Mo-Mo distance,
which controls the strength of AFM SE interactions between the $a_{1g}$ orbitals.
The latter competes with the FM interactions between the $e_g'$ orbitals, and dominates when
the Mo-Mo distance is small. In the frustrated pyrochlore lattice, this corresponds
to the transition into the SG state, though we cannot
specify precisely which mechanism freeze the random configuration of the Mo spins.
It could be the disorder of the Mo-Mo bondlengths proposed in Refs.~\onlinecite{Booth}
and \onlinecite{Keren}. Another possibility is the anisotropy of nearest-neighbor
magnetic interactions caused by the orbital ordering.\cite{comment.12}
It is also important to note that the Coulomb repulsion $U$
should be sufficiently strong in order to open the band gap
and suppress the FM double exchange interactions between the $e_g'$ orbitals.

  $\bullet$ The metal-insulator transition in these systems does not necessary coincide
with the FM-SG transitions. All metallic compounds are expected to be nearly collinear
ferromagnets, due to the double exchange mechanism. However, the inter-atomic
exchange coupling in the insulating state can be both ferromagnetic and antiferromagnetic,
depending on the Mo-Mo bondlength. The finding is qualitatively consistent with the
experimental data.\cite{Iikubo} The magnetic structure in the insulating
state is expected to be strongly non-collinear, due to the competition between the
isotropic exchange interactions and the single-ion anisotropies.

 $\bullet$ The large Hall conductivity can be explained by the combination
of three factors in the nearly collinear FM state: large SOI at the Mo sites, which
unquenches the orbital magnetization; strong enhancement of this effect by the on-site
Coulomb interaction; and the half-metallic electronic structure of the FM pyrochlores.
We also expect large magneto-optical effect in the same compounds.

  A possible way
to enhance these properties is to use heavy $5d$ elements, which have large
spin-orbit coupling. Unfortunately, all known $5d$ pyrochlores are non-magnetic, though
some of them are presumably close to the point of magnetic instability.\cite{Singh1}
Nevertheless, one can try to exploit an old idea known from the physics of intermetallic
alloys,\cite{PRB95} and replace part of the Mo sites
in $A_2$MoO$_7$ by the $5d$ elements. Then, the main question is whether the
remaining Mo sites will be
magnetic in the new environment or not. If they will, they will also magnetize the
$5d$ states at the neighboring sites, and in this way enhance the
non-reciprocal characteristics
of the considered pyrochlore compounds. As a purely model example, we have considered the ordered
Gd$_2$MoWO$_7$ alloy (using the lattice parameters of Gd$_2$Mo$_2$O$_7$). According
to the LSDA calculations, the system is expected to be ferromagnetic. The values of
(spin only)
magnetic moments at the Mo and W sites are 2.1 and 1.5 $\mu_B$, respectively.
The situation is highly unusual, because the examples when $4d$ and $5d$ elements
form a magnetic solution are rather rare. Although Gd$_2$MoWO$_7$ itself
may be rather unrealistic example from the technological point of view,
the idea to use Mo atoms in order
to magnetize the heavy $5d$ elements in the pyrochlore lattice seems to be
promising.

  In conclusion, the simple HF calculations allow to rationalize many
unusual properties of the pyrochlore compounds
$A_2$Mo$_2$O$_7$. It is also important that all these
properties can be explained by the universal value of $U$$\sim$$1.5$-$2.5$ eV, which
is of the same order of magnitude as the characteristic bandwidth for the
Mo($t_{2g}$) electrons. This will certainly place the considered compounds to the
direct proximity of the metal-insulator transition, meaning that the quantum
fluctuations beyond the mean-field HF approach will play an important
role too. This is certainly a very important direction for the
extension of the present analysis.

  Finally, we would like to make several comments on the comparison between our
calculations and the experimental magnetization\cite{Taguchi3,Yoshii,Ali}
and neutron diffraction\cite{Taguchi3,Yasui} data.

  (1) The experimental value of the Mo moment ($\sim$$1.1$-$1.3$ $\mu_B$) seems
to be smaller than $\sim$$1.4$ $\mu_B$ obtained in our mean-field calculations
for the nearly collinear FM state, even after taking into
account the orbital part ($\sim$$0.1$ $\mu_B$), which according to the third Hund's
rule is antiparallel to the spin one and partly compensate the latter. The small
difference is presumably related with the quantum fluctuations at the proximity
of the metal-insulator transition.\cite{comment.13}

  (2) The experimental data for Nd$_2$Mo$_2$O$_7$ and Sm$_2$Mo$_2$O$_7$
typically indicate at the existence of two antiferromagnetically coupled
sublattices. These sublattices are usually regarded as Mo and $A$ ones.
However, according to
the electronic structure point of view,
the {\it spin} coupling between these two sublattices should be
ferromagnetic,\cite{MoGd_exchange} being actually consistent
with the experimental data
for Gd$_2$Mo$_2$O$_7$.\cite{Ali} The controversies can be reconciled with
each other if the local magnetic moment at the Nd/Sm sites has a large orbital
counterpart, which is larger than the spin one. The AFM orientation of the spin
and orbital moments at the Nd/Sm sites is again the subject of the third
Hund's rule. Then, the magnetization and neutron diffraction experiments presumably
probe the AFM coupling between {\it spin} magnetic moments at the Mo sites and the
{\it orbital} magnetic moments at the Nd/Sm sites. Gd$^{3+}$ has no orbital moment,
and therefore the experimental data for this compound indicate at the conventional
FM coupling between two spin sublattices.
Then, the sharp decrease of the magnetization observed in Nd$_2$Mo$_2$O$_7$
at low $T$,\cite{Taguchi3}
which is frequently tempted to connect with the anomaly of the Hall
conductivity is due to the orbital magnetic structure, not the spin one.
The change of the {\it orbital} magnetic structure seems to be the right subject to
concentrate at when trying to explain, for example, the temperature dependence
of the AHE. In this respect, another interesting problem seems to be the behavior
of {\it spin and orbital} magnetic moments at the Mo sites, because already in the
ground state these two sublattices from mutually non-collinear magnetic structures.
Therefore, it is reasonable to expect very unconventional spin and orbital
dynamic at the elevated temperatures.

\section*{Acknowledgments}
I would like to thank Lars Nordstr\"{o}m for collaboration on early stages of this
project and kind hospitality during my stay at Uppsala University. I also benefited
from discussions with N.~Nagaosa, Y.~Tokura, K.~Terakura, and A.~B.~Granovski.

\newpage
\begin{table}
\caption{Structural parameters of $A_2$Mo$_2$O$_7$ ($A$$=$ Y, Nd, and Gd):
the cubic lattice parameter $a$ (in \AA),
positions of O $48f$ sites $[u,\frac{1}{8},\frac{1}{8}]$ (units of $a$),
the distances Mo-Mo and Mo-O (in \AA), and the angles Mo-O-Mo and O-Mo-O
(in degrees).}
\label{tab:structure}
\begin{ruledtabular}
\begin{tabular}{ccccccc}
compound          &   $a$      &    $u$    & $d_{\rm Mo-Mo}$ & $d_{\rm Mo-O}$ & $\angle$Mo-O-Mo & $\angle$O-Mo-O  \\
\tableline
Y$_2$Mo$_2$O$_7$\footnote{Ref.~\onlinecite{Reimers}, $a$ is from Ref.~\onlinecite{Gardner} (corresponding to 4K).}
& 10.21      &  0.33821  & 3.6098          & 2.0171         & 127.0           & 99.5            \\
Gd$_2$Mo$_2$O$_7$\footnote{Ref.~\onlinecite{Moritomo}.}
 & 10.3356(1) &  0.33158  & 3.6542          & 2.0123         & 130.4           & 97.2            \\
Nd$_2$Mo$_2$O$_7$\footnote{Ref.~\onlinecite{Moritomo}.}
 & 10.4836(2) &  0.32977  & 3.7065          & 2.0332         & 131.4           & 96.6            \\
\end{tabular}
\end{ruledtabular}
\end{table}

\newpage

\begin{thebibliography}{10}

\bibitem{Taguchi3}
Y.~Taguchi, Y.~Oohara, H.~Yoshizawa, N.~Nagaosa, and Y.~Tokura,
Science~{\bf 291}, 2573 (2001).

\bibitem{Katsufuji}
T.~Katsufuji, H.~Y.~Hwang, and S-W.~Cheong,
Phys.~Rev.~Lett.~{\bf 84}, 1998 (2000).

\bibitem{Moritomo}
Y.~Moritomo, Sh.~Xu, A.~Machida, T.~Katsufuji, E.~Nishibori, M.~Takata,
M.~Sakata, and S-W.~Cheong,
Phys.~Rev.~B~{\bf 63}, 144425 (2001).

\bibitem{Taguchi2}
Y.~Taguchi, K.~Ohgushi, and Y.~Tokura,
Phys.~Rev.~B~{\bf 65}, 115102 (2002).

\bibitem{Iikubo}
S.~Iikubo, S.~Yoshii, T.~Kageyama, K.~Oda, Y.~Kondo, K.~Murata, and M.~Sato,
J.~Phys.~Soc.~Jpn.~{\bf 70}, 212 (2001).

\bibitem{Reimers}
J.~N.~Reimers, J.~E.~Greedan, and M.~Sato,
J.~Solid~State~Chem.~{\bf 72}, 390 (1988).

\bibitem{Gardner}
J.~S.~Gardner, B.~D.~Gaulin, S.-H.~Lee, C.~Broholm, N.~P.~Raju, and
J.~E.~Greedan,
Phys.~Rev.~Lett.~{\bf 83}, 211 (1999).

\bibitem{Reimers2}
J.~N.~Reimers, Phys.~Rev.~B~{\bf 45}, 7287 (1992).

\bibitem{Booth}
C.~H.~Booth, J.~S.~Gardner, G.~H.~Kwei, R.~H.~Heffner, F.~Bridges, and
M.~A.~Subramanian, Phys.~Rev.~B~{\bf 62}, R755 (2000).

\bibitem{Keren}
A.~Keren and J.~S.~Gardner,
Phys.~Rev.~Lett.~{\bf 87}, 177201 (2001).

\bibitem{Moriya}
T.~Moriya, Phys.~Rev.~{\bf 117}, 635 (1960).

\bibitem{AHE_manganites}
J.~Ye, Y.~B.~Kim, A.~J.~Millis, B.~I.~Shraiman, P.~Majumdar, and Z.~Te\v{s}anovish,
Phys.~Rev.~Lett.~{\bf 83}, 3737 (1999).
Y.~Lyanda-Geller, S.~H.~Chun, M.~B.~Salamon, P.~M.~Goldbart, P.~D.~Han,
Y.~Tomioka, A.~Asamitsu, and Y.~Tokura, Phys.~Rev.~B~{\bf 63}, 184426 (2001).

\bibitem{KL}
R.~Karplus and J.~M.~Luttinger, Phys.~Rev.~{\bf 95}, 1154 (1954).

\bibitem{AHE_history}
J.~Smith, Physica~{\bf 21}, 877 (1955);
J.~Smith, {\it ibid.}~{\bf 24}, 39 (1958);
J.~M.~Luttinger, Phys.~Rev.~{\bf 112}, 739 (1958);
J.~Kondo, Prog.~Theor.~Phys.~{\bf 27}, 772 (1962);
F.~E.~Maranzana, Phys.~Rev.~{\bf 160}, 421 (1967);
L.~Berger, Phys.~Rev.~B~{\bf 2}, 4559 (1970);
L.~Berger, {\it ibid.}~{\bf 5}, 1862 (1972);
S.~K.~Lyo and T.~Holstein, Phys.~Rev.~Lett.~{\bf 29}, 423 (1972);
J.~Smith, Phys.~Rev.~B~{\bf 17}, 1450 (1978).

\bibitem{comment.14}
We would not like to go here very deeply into all details of these
controversies. Briefly, Karplus and Luttinger in their pioneering work\cite{KL}
considered the contributions proportional to $\partial f/\partial \varepsilon$,
where $f$ is the Fermi distribution function. These contributions have the form
of intra-band transitions in the conductivity tensor depicted by Eq.~(\ref{eqn:sintra}).
All of them vanish in the periodic crystal, as it was correctly pointed out by
Smith.\cite{AHE_history} However, if the product of two momentum operators,
$\pi^\alpha_{nn'} \pi^\beta_{n'n}$, has an imaginary part, one can expect another
contribution to $\sigma_H$ coming from Eq.~(\ref{eqn:sinter}). This contribution
is caused by the inter-band transitions, and in this sense the considered mechanism
has the same origin as the magneto-optical effect. This term was intensively studied
recently by Nagaosa {\it et~al.} in the contexy of the spin-chirality
problem.\cite{Nagaosa} See also comments in Ref.~\onlinecite{comment.10}.

\bibitem{Kondorskii}
A.~B.~Granovskiy and Ye.~I.~Kondorskiy,
Fiz.~Metal.~Metalloved.~{\bf 39}, 718 (1975)
[English transl.: The~Physics~of~Metals~and~Metallography~{\bf 39}, No.4, p.36 (1975)];
Ye.~I.~Kondorskiy, A.~V.~Vedyayev, and A.~B.~Granovskiy,
{\it ibid.}~{\bf 40}, 455 (1975)
[English transl.: The~Physics~of~Metals~and~Metallography~{\bf 40}, No.3, p.1 (1975)];
N.~Ryzhanova, A.~Vedyayev, A.~Cr\'{e}pieux, and C.~Lacroix,
Phys.~Rev.~B~{\bf 57}, 2943 (1998).

\bibitem{comment.10}
Using expression (\ref{eqn:sinter}) for the conductivity tensor, it is easy to show
that $\sigma^*_{\alpha \beta}(\omega)$$=$$\sigma_{\alpha \beta}($$-$$\omega)$.
Therefore, ${\rm Im} \sigma_{\alpha \beta}(0)$$=$$0$, whereas
$\sigma_H$$\equiv$${\rm Re} \sigma_{\alpha \beta}(0)$ can be finite (the author thank
N.~Nagaosa for pointing out this property). On the other hand, the Onsager constraint
$\sigma_{\alpha \beta}(\omega,\mu)$$=$$\sigma_{\beta \alpha}(\omega,$$-$$\mu)$
($\mu$ being the net magnetization) requires $\sigma_H$ to be an even
function of $\mu$.\cite{LandauLifshitz} Strictly speaking, the effect is masked by
the diagonal elements of the conductivity tensor, and since
${\rm Re}\sigma_{xx}(0)$$\rightarrow$$\infty$,
the resistivity
$\rho_{xy}$$=$$-$${\rm Re}\sigma_{xy}(0)/{\rm Re}\sigma^2_{xx}(0)$, which is
the measured quantity, will vanish anyway
in the perfectly periodic metallic state. In this
sence, additional scattering mechanisms\cite{AHE_history,Kondorskii}
are indispensable in order to have finite
${\rm Re}\sigma_{xx}(0)$ and $\rho_{xy}$. However, it does not mean that
${\rm Re}\sigma_{xy}(0)$ itself will vanish in the periodic state. In our work
we will consider such, perhaps somewhat idealistic, situation.

\bibitem{Nagaosa}
K.~Ohgushi, S.~Murakami, and N.~Nagaosa,
Phys.~Rev.~B~{\bf 62}, R6065 (2000);
R.~Shindou and N.~Nagaosa,
Phys.~Rev.~Lett.~{\bf 87}, 116801 (2001);
M.~Onoda and N.~Nagaosa,
J.~Phys.~Soc.~Jpn.~{\bf 71}, 19 (2001).

\bibitem{Dzyaloshinskii}
I.~E.~Dzyaloshinskii, Phys.~Lett.~A~{\bf 155}, 62 (1991);
I.~Dzyaloshinskii and E.~V.~Papamichail,
Phys.~Rev.~Lett.~{\bf 75}, 3004 (1995).

\bibitem{PRL98}
I.~V.~Solovyev, A.~I.~Liechtenstein, and K.~Terakura,
Phys.~Rev.~Lett.~{\bf 80}, 5758 (1998).

\bibitem{Irkhin}
V.~Yu.~Irkhin and M.~I.~Katsnel'son,
Usp.~Fiz.~Nauk~{\bf 164}, 705 (1994)
[English transl.: Physics--Uspekhi~{\bf 37}, 659 (1994)].

\bibitem{Yoshii}
S.~Yoshii, S.~Iikubo, T.~Kageyama, K.~Oda, Y.~Kondo, K.~Murata, and M.~Sato,
J.~Phys.~Soc.~Jpn.~{\bf 69}, 3777 (2002).

\bibitem{Taguchi}
Y.~Taguchi and Y.~Tokura, Phys.~Rev.~B~{\bf 60}, 10280 (1999).

\bibitem{Singh}
D.~J.~Singh, J.~Appl.~Phys.~{\bf 79}, 4818 (1996).

\bibitem{comment.3}
In the local coordinate frame, the $a_{1g}$ and two $e_g'$ orbitals
have the following form: $|a_{1g}\rangle$$=$$\frac{1}{\sqrt{3}}
(|xy\rangle$$+$$|yz\rangle$$+$$|zx\rangle)$,
$|e_{g1}'\rangle$$=$$\frac{1}{\sqrt{2}}(|yz\rangle$$-$$|zx\rangle)$, and
$|e_{g2}'\rangle$$=$$\frac{1}{\sqrt{6}}($$-$$2|xy\rangle$$+$$|yz\rangle$$+$$
|zx\rangle)$.

\bibitem{LiV2O4}
In this sence, the physics is similar to the spinel compounds:
V.~I.~Anisimov, M.~A.~Korotin, M.~Z\"{o}lfl, T.~Pruschke, K.~Le~Hur, and
T.~M.~Rice, Phys.~Rev.~Lett.~{\bf 80}, 5758 (1998).
Note, however, that the oxygen coordination is
very different in the spinel and pyrochlore structure.

\bibitem{PRB94}
I.~V.~Solovyev, P.~H.~Dederichs, and V.~I.~Anisimov,
Phys.~Rev.~B~{\bf 50}, 16861 (1994).

\bibitem{PRB96}
I.~Solovyev, N.~Hamada, and K.~Terakura,
Phys.~Rev.~B~{\bf 53}, 7158 (1996).

\bibitem{LMTO}
O.~K.~Andersen, Phys.~Rev.~B~{\bf 12}, 3060 (1975);
O.~Gunnarsson, O.~Jepsen, and O.~K.~Andersen,
{\it ibid.}~{\bf 27}, 7144 (1983).

\bibitem{Hamada}
N.~Hamada, H.~Sawada, and K.~Terakura (unpublished).

\bibitem{Lars}
L.~Nordstr\"{o}m, to be published.

\bibitem{Singh1}
D.~J.~Singh, P.~Blaha, K.~Schwarz, and J.~O.~Sofo,
Phys.~Rev.~B~{\bf 65}, 155109 (2002).

\bibitem{PRB98}
I.~V.~Solovyev and K.~Terakura, Phys.~Rev.~B~{\bf 58}, 15496 (1998).

\bibitem{KugelKhomskii}
K.~I.~Kugel and D.~I.~Khomskii,
Usp.~Fiz.~Nauk~{\bf 136}, 621 (1982)
[Sov.~Phys.~Usp.~{\bf 25}, 231 (1982)].

\bibitem{comment.4}
By considering the {\it nn} hoppings between Mo($a_{1g}$) orbitals
in the pyrochlore lattice, $t_{a_{1g}}$, it is easy to show that the
electronic structure for the $a_{1g}$ states will consist of three
bands (in units of $t_{a_{1g}}$): $\varepsilon_{{\bf k} \pm}$$=$$-$$2$$\pm
\sqrt{1+s_{\bf k}}$, the so-called bonding and antibonding bands, and the
doubly-degenerate non-bonding
band $\varepsilon_{{\bf k} 0}$$=$$2$, where
$s_{\bf k}$$=$$\cos \frac{k_xa}{2} \cos \frac{k_ya}{2}$$+
$$\cos \frac{k_ya}{2} \cos \frac{k_za}{2}$$+$$\cos \frac{k_za}{2} \cos \frac{k_xa}{2}$.

\bibitem{comment.5}
In fact, there is a non-spherical (proportional to the
Coulomb parameter $B$) interaction
beween
occupied $e_g'$ and $a_{1g}$ orbitals, which will deforms the density of the
$a_{1g}$ electrons. This interaction has an appreciable effect in the AFM
state. For example, for Y$_2$Mo$_2$O$_7$ and $U$$=$$3.0$ eV, the occupied
$a_{1g}$ orbital at the site 1 has the form
$| a_{1g} \rangle$$=$$\alpha|xy\rangle$$+$$\beta|yz\rangle$$+$$\beta|zx\rangle$,
where $\alpha$$\simeq$$0.629 \ne \beta$$\simeq$$0.549$. This will
also lead to a small (less than 1 meV) anisotropy of the
$a_{1g}$-$a_{1g}$ superexchange interactions.

\bibitem{Liechtenstein}
A.~I.~Liechtenstein, M.~I.~Katsnelson, V.~P.~Antropov, and
V.~A.~Gubanov, J.~Magn.~Magn.~Matter.~{\bf 67}, 65 (1987).

\bibitem{comment.2}
Note that $J_{\bm{\tau} \bm{\tau}'}$ defined by
Eq.~(\ref{eqn:exchange}), which is basically the second derivative
of the total energy,\cite{Liechtenstein} characterizes the
{\it local stability} of the magnetic state, and provides
a {\it complementary} information to the standard total energy calculations.
Generally, the parameters $J_{\bm{\tau} \bm{\tau}'}$
depend on the magnetic state in which they are calculated and there is no
simple relation between $J_{\bm{\tau} \bm{\tau}'}$
and the total energy.
On the other hand, the
total energy difference calculated for a
limited number of magnetic states does not necessary guarantee that
these states are locally stable to be
the local minima
of the total energy.

\bibitem{MoGd_exchange}
The Gd($4f$) core spin polarizes the valence, mainly Gd($5d$), electrons,
which can interact with the Mo($4d$) electrons. According to our estimates
based on Eq.~\ref{eqn:exchange}, the interaction
$J_{5d-4d}$ is ferromagnetic and is of the order of 2 meV (in LSDA, and treating
the Gd($4f$) states as the core states), in a good agreement with
the experimantal data.\cite{Ali} Similar conclusion can be obtained
from the analysis of the total energy difference, assuming the FM and AFM coupling
between Mo and Gd spins. Results of these calculations have very transparent
expanation, because Gd$_2$Mo$_2$O$_7$ has half-metallic electronic structure
and the FM coupling between Mo and Gd spins results in the additional energy gain
for the $\uparrow$-spin Mo($t_{2g}$) electrons (see Fig.~\ref{fig.DOSLSDA}).
For comparison, the direct Gd($5d$)-Gd($5d$) interaction is much smaller
($\sim$$0.03$ meV, which is formally beyond the accuracy verge of the present
calculations).

\bibitem{Springer02}
I.~V.~Solovyev and K.~Terakura, to be published in
{\it Electronic Structure and Magnetism of Complex Materials},
ed. by D.~J.~Singh (Springer-Verlag, Berlin, 2002).

\bibitem{comment.1}
Presumably, other factors can also play some role in explaining the large
difference of $a_{1g}$-$a_{1g}$ interactions obtained for Y$_2$Mo$_2$O$_7$ and
Nd$_2$Mo$_2$O$_7$.
The standard SE interaction is proportional to the (minus) second moment
of the density of states, that actually explains the AFM sign of the
coupling.\cite{Springer02}
The $\uparrow$-spin $a_{1g}$ bandwidths in Y$_2$Mo$_2$O$_7$ and Nd$_2$Mo$_2$O$_7$
differ by factor two (Fig.~\ref{fig.DOS30}), that should lead to the factor four
difference for corresponding SE interactions. However, the $a_{1g}$ states
are not fully isolated and interact with some other states,
so that the expression
for the SE interaction should be modified as
$J^S \propto {\rm Im} \int_{-\infty}^{\varepsilon_F} d\varepsilon
(\varepsilon-c_\uparrow)(\varepsilon-c_\downarrow){\rm Tr}_L
{\cal G}^\uparrow_{\bm{\tau} \bm{\tau}}(\varepsilon)$, where
$c_{\uparrow,\downarrow}$ is the shift of the $a_{1g}$ states due to the
hybridization with other states. This shift can be
different for the $\uparrow$- and $\downarrow$-spin electrons
and also depend on the material,
that presumably
explains an additional difference between Y$_2$Mo$_2$O$_7$ and Nd$_2$Mo$_2$O$_7$.

\bibitem{comment.7}
Note that at least in the mean-field approximation for the one-orbital model, the effects of the
inter-site Coulomb interactions, which are responsible for the charge disproportionation,
can be included as a renormalization of the local (on-site) interaction, in the
spirit of model HF calculations performed in this work:
I.~V.~Solovyev, Phys.~Rev.~B~{\bf 60}, 8550 (1999).

\bibitem{Bruno}
P.~Bruno, Phys.~Rev.~B~{\bf 39}, 865 (1989).

\bibitem{JMMM1998}
I.~V.~Solovyev, A.~I.~Liechtenstein, and K.~Terakura,
J.~Magn.~Magn.~Matter.~{\bf 185}, 118 (1998).

\bibitem{comment.6}
The conclusion can be easily verified by considering the spin Hamiltonian
${\cal H}_S$$=$$-$$J\sum_{\langle \bm{\tau} \bm{\tau}' \rangle }
({\bf e}_{\bm{\tau}},{\bf e}_{\bm{\tau}'})$$+$$\sum_{\bm{\tau}}
({\bf e}_{\bm{\tau}},\widehat{\mathbb{D}}_{\bm{\tau}} {\bf e}_{\bm{\tau}} )$,
consisting of the {\it nn} exchange and anisotropy parts.
$\mathbb{D}_1$$=$$d \| \frac{2}{3}\delta_{\alpha \beta} - 1 \|$
is the single-ion anisotropy tensor at the site 1.
$\widehat{\mathbb{D}}_2$, $\widehat{\mathbb{D}}_3$, and
$\widehat{\mathbb{D}}_4$ are obtained by the
180$^\circ$ rotations of $\widehat{\mathbb{D}}_1$ around
$x$, $y$, and $z$, respectively. Then if $J$$>$$0$ (the FM coupling) and $d$$>$$0$
(the easy axes scenario), the ''umbrella'' structure has lower energy in
comparison with the ''two-in, two-out'' structure due to the
additional energy gain coming from the {\it nn} exchange interactions.

\bibitem{Sandratskii}
L.~M.~Sandratskii and J.~K\"{u}bler,
Phys.~Rev.~Lett.~{\bf 76}, 4963 (1996).

\bibitem{PRB97}
I.~V.~Solovyev,
Phys.~Rev.~B~{\bf 55}, 8060 (1997).

\bibitem{comment.8}
Note that $\sigma_H$$\propto$$\delta_G^{-2}$, where $\delta_G$ is the
band gap. In addition, the non-collinear alignment reduces the net
magnetization and the conventional contribution to the AHE caused by
the SOI.\cite{KL}

\bibitem{method_optica}
Yu.~A.~Uspenski, E.~G.~Maksimov, S.~N.~Rashkeev, and I.~I.~Mazin,
Z.~Phys.~B~{\bf 53}, 263 (1983).

\bibitem{method_moptica}
C.~S.~Wang and J.~Callaway,
Phys.~Rev.~B~{\bf 9}, 4897 (1974);
Yu.~A.~Uspenski and S.~V.~Khalilov,
Zh.~Eksp.~Teor.~Fiz.~{\bf 95}, 1022 (1989)
[English transl.: Sov.~Phys.~JETP~{\bf 68}, 588 (1989)];
P.~M.~Oppeneer, T.~Maurer, J.~Sticht, and J.~K\"{u}bler,
Phys.~Rev.~B~{\bf 45}, 10924 (1992).

\bibitem{Uspenskii}
R.~Pittini, J.~Schoenes, O.~Vogt, and P.~Wachter,
Phys.~Rev.~Lett.~{\bf 77}, 944 (1996).
Yu.~A.~Uspenskii and B.~N.~Harmon,
Appl.~Phys.~Lett.~{\bf 74}, 1618 (1999).

\bibitem{comment.9}
In practical calculations, we replace the wave functions $| \widetilde{n {\bf k}} \rangle$
by $(1$$-$$\widehat{P}_{a_{1g}})| \widetilde{n {\bf k}} \rangle$, where
$\widehat{P}_{a_{1g}}$ is the projection operator onto the $a_{1g}$ states.

\bibitem{TokuraNagaosa}
Y.~Tokura and N.~Nagaosa,
Science~{\bf 288}, 462 (2000).

\bibitem{comment.11}
Another possibility could be the sample purity. However, according to the experimental estimates
the sample quality seems to be good (for example in Y$_2$Mo$_2$O$_7$
the concentration of anti-site defects and deviation from the nominal oxygen stoichiometry
was estimated as 4\% and 1\%, respectively).\cite{Keren}

\bibitem{comment.12}
Strictly speaking, two mechanisms are closely related because the orbital
ordering will produce the lattice distortion and vice versa.

\bibitem{PRB95}
I.~V.~Solovyev, P.~H.~Dederichs, and I.~Mertig,
Phys.~Rev.~B~{\bf 52}, 13419 (1995).

\bibitem{Ali}
N.~Ali, M.~P.~Hill, S.~Labroo, and J.~E.~Greedan,
J.~Solid~State~Chem.~{\bf 83}, 178 (1989).

\bibitem{Yasui}
Y.~Yasui, Y.~Kondo, M.~Kanada, M.~Ito, H.~Harashina, M.~Sato, and K.~Kakurai,
J.~Phys.~Soc.~Jpn.~{\bf 70}, 284 (2001).

\bibitem{comment.13}
Another indication that the FM pyrochlores may not be quite conventional
metals is the relatively high resistivity
($\sim$$10^{-3}\Omega {\rm cm}$).\cite{Taguchi2}

\bibitem{LandauLifshitz}
D.~L.~Landau and E.~M.~Lifshitz, {\it Electrodymanics of Condensed Media}
(Pergamon, New York, 1960).

\end{thebibliography}
\end{document}